\documentclass[10pt,twocolumn,twoside]{IEEEtran}

\usepackage{amssymb}
\usepackage{graphicx}

\usepackage{url}

\usepackage[normalem]{ulem}
\usepackage{booktabs} 
\usepackage{amsmath,amsfonts,amssymb}
\usepackage{subfig}

\newcommand*{\Fitcol}[1]{\resizebox{0.45\textwidth}{!}{$#1$}}
\newcommand{\bt}[1]{\mbox{$\bf #1$}}

\usepackage{xifthen}

\newcommand{\toremove}[2]{#2}

\newcommand{\F}[0]{\mathbf{\Lambda_F}}
\newcommand{\f}[0]{\mathbf{f}}

\newcommand{\pest}[2]{\hat{p}_{#1,#2}}

\newcommand{\Prob}[2]{\ensuremath{P_{#1,#2}}}

\newcommand{\identity}[1]{\ensuremath{\mathbf{I}_{#1}}}
\newcommand{\probsimplex}[0]{\ensuremath{\mathcal{P}}}

\newcommand{\nusers}[0]{\ensuremath{N}}
\newcommand{\thre}[0]{\ensuremath{t}}

\newcommand{\send}[2]{\ensuremath{x_{#1}^{#2}}}
\newcommand{\sendv}[1]{\ensuremath{\mathbf{x}^{#1}}}
\newcommand{\sendm}[0]{\ensuremath{\mathbf{U}}}
\newcommand{\sendr}[2]{\ensuremath{X_{#1}^{#2}}}

\newcommand{\recv}[2]{\ensuremath{y_{#1}^{#2}}}
\newcommand{\recvv}[2]{%
  \ifthenelse{\isempty{#2}}%
    {\ensuremath{\mathbf{y}_{#1}}}
    {\ensuremath{\mathbf{y}_{#1}^{#2}}}
}

\newcommand{\recvr}[2]{\ensuremath{Y_{#1}^{#2}}}
\newcommand{\recvvr}[2]{\ensuremath{\mathbf{Y}_{#1}^{#2}}}

\newcommand{\Hm}[0]{\ensuremath{\mathbf{H}}}

\newcommand{\Hmpoolest}[0]{\ensuremath{\hat{\mathbf{H}}_s}}
\newcommand{\recvall}[0]{\ensuremath{\mathbf{y}}}
\newcommand{\recvallr}[0]{\ensuremath{\mathbf{Y}}}
\newcommand{\probsall}[0]{\ensuremath{\mathbf{p}}}
\newcommand{\probsallreal}[0]{\ensuremath{\mathbf{p}}}
\newcommand{\probsallest}[0]{\ensuremath{\hat{\mathbf{p}}}}
\newcommand{\probsallcand}[0]{\ensuremath{\mathbf{p}}} 

\newcommand{\sendpool}[2]{%
  \ifthenelse{\isempty{#1}}%
    {x_s^{#2}}
    {x_{s,#1}^{#2}}
}

\newcommand{\sendpoolm}[0]{\ensuremath{\mathbf{U}_s}}

\newcommand{\sendpoolr}[2]{%
  \ifthenelse{\isempty{#1}}%
    {X_s^{#2}}
    {X_{s,#1}^{#2}}
}
\newcommand{\sendpoolvr}[1]{\ensuremath{\mathbf{X}_s^{#1}}}
\newcommand{\sendpoolmr}[0]{\ensuremath{\mathbf{U}_s}}

\newcommand{\sendpoolest}[2]{%
  \ifthenelse{\isempty{#1}}%
    {\hat{x}_s^{#2}}
    {\hat{x}_{s,#1}^{#2}}
}

\newcommand{\sendpoolmest}[0]{\ensuremath{\hat{\mathbf{U}}_s}}

\newcommand{\sendpoolearlier}[2]{
  \ifthenelse{\isempty{#1}}%
    {X_e^{#2}}
    {X_{e,#1}^{#2}}
}
\newcommand{\sendpoolnoise}[2]{\ensuremath{N_{#1}^{#2}}}

\newcommand{\B}[0]{\ensuremath{\mathbf{B}}}
\newcommand{\Nzero}[0]{\ensuremath{\mathbf{N}_0}}

\newcommand{\prob}[2]{\ensuremath{p_{#1,#2}}}
\newcommand{\sendprof}[1]{\ensuremath{\mathbf{q}_{#1}}}
\newcommand{\recvprof}[1]{\ensuremath{\mathbf{p}_{#1}}}

\newcommand{\transprobest}[0]{\hat{\ensuremath{\mathbf{P}}}}
\newcommand{\recvprofest}[1]{\ensuremath{\hat{\mathbf{p}}_{#1}}}
\newcommand{\sendprofest}[1]{\ensuremath{\hat{\mathbf{q}}_{#1}}}

\newcommand{\probest}[2]{\ensuremath{\hat{p}_{#1,#2}}}
\newcommand{\sendfreq}[1]{\ensuremath{f_{#1}}}
\newcommand{\sendfreqest}[1]{\ensuremath{\hat{f}_{#1}}}
\newcommand{\sendfreqm}[0]{\F}

\newcommand{\binvar}[1]{\ensuremath{s_{j,#1}}}
\newcommand{\binvarm}[0]{\ensuremath{\mathbf{S}_j}}
\newcommand{\uniformi}[1]{\ensuremath{\mu_{#1}}}
\newcommand{\meanuniformi}[0]{\ensuremath{\bar{\mu}}}
\newcommand{\nfriends}[0]{\ensuremath{n_f}}

\newcommand{\MSEi}[0]{\ensuremath{\mbox{MSE}_i}}
\newcommand{\MSEp}[0]{\ensuremath{\mbox{MSE}_{p}}}

\newcommand{\autocorr}[0]{\ensuremath{\mathbf{R}_{x}}}
\newcommand{\autocorrpool}[0]{\ensuremath{\mathbf{R}_{xs}}}

\newcommand{\recvprofm}[1]{\ensuremath{\mathbf{P}_{#1}}}

\newcommand{\alphaq}[0]{\ensuremath{\alpha_q}}

\newcommand{\stepsize}[0]{\ensuremath{\tau}} 

\begin{document}

\title{A Least Squares Approach to the\\ Static Traffic Analysis of High-Latency\\ Anonymous Communication Systems}

\author{Fernando~P\'erez-Gonz\'alez*,~\IEEEmembership{Senior Member,~IEEE,}~Carmela~Troncoso and~Simon~Oya%
\thanks{F. P\'erez-Gonz\'alez and S. Oya are with the Signal Theory and Communications Dept., University of Vigo. F. P\'erez-Gonz\'alez and C. Troncoso are with Gradiant (Galician R\&D Center in Advanced Telecommunications). }%
\thanks{This work was supported in part by the Spanish Government and the European Regional Development Fund under project TACTICA, in part by the Galician Regional Government under project Consolidation of Research Units GRC2013/009, and by the EU 7th Framework Programme (FP7/2007-2013) under grant agreements 610613 (PRIPARE) and 285901 (LIFTGATE).}}

\maketitle

\begin{abstract}
Mixes, relaying routers that hide the relation between incoming and outgoing messages, are the main building block of high-latency anonymous communication networks. A number of so-called disclosure attacks have been proposed to effectively de-anonymize traffic sent through these channels. Yet, the dependence of their success on the system parameters is not well-understood. We propose the Least Squares Disclosure Attack (LSDA), in which user profiles are estimated by solving a least squares problem. We show that LSDA is not only suitable for the analysis of threshold mixes, but can be easily extended to attack pool mixes. Furthermore, contrary to previous heuristic-based attacks, our approach allows us to analytically derive expressions that characterize the profiling error of LSDA with respect to the system parameters. We empirically demonstrate that LSDA recovers  users' profiles with greater accuracy than its statistical predecessors and verify that our analysis closely predicts actual performance.
\end{abstract}


\IEEEpeerreviewmaketitle

\section{Introduction}
%
%
%
%

%
%

\IEEEPARstart{C}{ommunication} confidentiality is traditionally achieved through cryptographic means. This protection, however, usually targets communication content and leaves network information accessible to potential adversaries. These traffic data, such as the identities of the participants in the communication (e.g. IP addresses), their location, or the amount and timing of data transferred, can be exploited by a passive observer to infer sensitive private information about the communication.

A well-known countermeasure against traffic analysis for high-latency anonymous communications, i.e., communications that tolerate delay (e.g., e-mail), is the use of mix networks~\cite{DDS09,DDM03,EY10,MCPS03}. Mixes prevent an observer from tracking communications by hiding the correspondence between inputs and outputs~\cite{Chaum81}. However, it is known that persistent and repeated communication patterns can be uncovered by means of a disclosure attack~\cite{AK03,Danezis03,DDT07,DanS04,DT09,KP04,MD04,MW10,PWK11,TGPV08}. In short, these attacks infer Alice's likely set of contacts, also known as her user profile, by observing the sets of possible receivers for each message Alice sends and processing this information.

The variants of the disclosure attack differ on the technique used to infer user profiles from the observed communications. Even though all of them have been proven effective at the time of de-anonymization/profiling, their heuristic nature and/or their complexity hinders the analysis of how system parameters influence their success. Furthermore, the great majority of attacks have only been evaluated against simple threshold mixes, where mixing occurs only between messages in a given round, and only the Statistical Disclosure Attack has been extended to attack pool mixes, in which messages can be delayed for more than one round~\cite{DanS04}.

In this paper we propose a profiling approach based on solving a least squares problem, the Least Squares Disclosure Attack (LSDA). This approach ensures that the error between the actual number of messages each user receives from the mix and a prediction based on the messages sent to the mix is minimized. We show that LSDA is an efficient estimator when users' behavior is static, i.e., the error incurred when estimating the user profiles asymptotically tends to zero as the number of observed mix rounds grows when user behavior does not change within the observation window, and that it is suitable to attack anonymous communication through both threshold and pool mixes. In particular, in this paper we consider threshold binomial pool mixes, in which messages are individually selected to stay in the mix or to be sent to their receiver according to a binomial distribution. We note, however, that the choice of this mix is arbitrary and our approach can be adapted to many other probabilistic mixing strategies~\cite{WIFS12}.

We provide two variants of LSDA: a very efficient unconstrained profile estimator that outputs user profiles that may contain negative probabilities (usually corresponding to receivers that are not contacts of the target user) and a slower constrained version that further minimizes the error by ensuring that the output profiles are well-defined. We show through simulations that the latter indeed minimizes the mean squared error with respect to heuristic disclosure attack variants~\cite{Danezis03,DanS04,TGPV08} although it performs slightly worse than the Bayesian approach~\cite{DT09} in the simple threshold mix scenario. In the pool mix scenario, however, applying the Bayesian inference techniques is computationally unfeasible and LSDA emerges as the best option to infer the user's behavior.

A remarkable feature of the least squares approach is that it allows for the derivation of analytical expressions that describe the evolution of the profiling error with the system parameters. This is a key property, as it permits designers to choose system parameters that provide a certain level of protection without needing to run simulations. We empirically validate our results, showing that our formulas 
reliably predict the evolution of LSDA's error as the parameters of the system change when users' behavior is static (i.e., time-invariant), and show the usefulness of our methodology beyond this assumptions using real data.

We note that previous works evaluated the attacks either from mostly a de-anonymization of individual messages perspective (e.g.,~\cite{DT09,TGPV08}) or from the point of view of the number of rounds necessary to identify a percentage of Alice's recipients (e.g.,~\cite{MW10,MD04,PWK11}). In this work we are interested in the accuracy with which the adversary can infer the sender profile of Alice, i.e., we not only seek to identify Alice's messages receivers, but also to estimate the probability that Alice sends a message to them.

The rest of the paper is organized as follows: in the next section we revisit previous work on disclosure attacks and we describe our system and adversarial models in Sect.~\ref{sec:sysmodel}. We introduce the least squares approach to disclosure applied to threshold mixes in Sect.~\ref{sec:LSDAthre}, and extend it to account for the pool mix in Sect.~\ref{sec:LSDApool}. In both sections we derive equations that characterize LSDA's error with respect to the system parameters which we validate in Sect.~\ref{sec:eval}. We discuss the limitations of our model and explain how to extend our analysis to more realistic scenarios in Sect.~\ref{sec:discussion}, and we conclude in Sect.~\ref{sec:conclusion}. 
\section{An Overview of Disclosure Attacks}

The first Disclosure Attack~\cite{AK03,KAP02} relies on graph theory to uncover the recipient set of a target user Alice. It identifies the set of Alice's contacts by seeking mutually disjoint sets of receivers among the recipient anonymity sets of the messages sent by Alice. The main drawback of this approach is that it is equivalent to solving a constraint satisfaction problem which is well-known to be NP-complete.

The subfamily of Hitting Set Attacks~\cite{KP04,PWK11} speeds up the search for Alice's messages recipients by restricting the search to unique minimal hitting sets. Pham et al. studied the relationship between the number of observed rounds to uniquely identify the set of receivers and the parameters of the system~\cite{PWK11}. This evaluation is similar to our work in spirit, but it focuses on attacks that unambiguously identify recipient sets while we deal with statistical attacks that only provide an estimation of such sets as the ones discussed below.

The Statistical Disclosure Attack (SDA), originally proposed by Danezis~\cite{DanS04}, and its sequels~\cite{DDT07,MD04,MW10}, estimate Alice's sending profile by averaging the probability distributions describing the recipient anonymity set~\cite{SD02} of her messages.
Mathewson and Dingledine improved Danezis' SDA by extending it to a more general scenario and to more complex mixing algorithms~\cite{MD04}. This improved version of the attack \toremove{(which we denote SDA-MD)}{} is able to isolate Alice's behavior by first estimating the behavior of all the remaining users, employing those observations where Alice has not participated.

Troncoso et al.~proposed in~\cite{TGPV08} two attacks: the Perfect Matching Disclosure Attack (PMDA) and the Normalized Statistical Disclosure Attack (NSDA). These attacks exploit the fact that the relationship between sent and received messages in a round must be one-to-one to improve the accuracy of the estimated profiles. PMDA accounts for this interdependency by searching for perfect matchings in the underlying bipartite graph representing a mix round, while NSDA normalizes the adjacency matrix representing this graph. The recipient anonymity set of a message is built based on the result of this assignment, instead of assigning uniform probabilities among all recipients as SDA does.

Last, Danezis and Troncoso propose to use Bayesian sampling techniques to co-infer users' profiles and de-anonymize messages~\cite{DT09}. The Bayesian approach outputs samples from the distribution of all possible sending profiles, which in turn allows to infer reliable error estimates. However, Vida requires the adversary to repeatedly seek for perfect matchings, increasing the computational requirements of the attack.

From all of the aforementioned attacks, only SDA has been extended to take into account pool mixes. The fact that a message can be delayed multiple rounds before being forwarded to its recipient largely increases the set of possible receivers of each message. This makes extending the Disclosure and Hitting Set attacks~\cite{AK03,KP04} a non trivial task and finding correspondences between incoming and outgoing messages, such as in PMDA, NSDA~\cite{TGPV08} and Vida~\cite{DT09}, computationally unfeasible. We will show that our least squares approach can be \toremove{easily}{} adapted to the pool mix probabilistic behavior without increasing significantly the computational resources needed.

\section{System and Adversary Model}
\label{sec:sysmodel}

In this section we describe our model of an anonymous communication system and introduce the notation we use throughout the paper, which we summarize in Table~\ref{tab:notation}. Capital letters denote random variables and lowercase letters denote realizations. Vectors are represented by boldface characters; thus, $\bt x=[x_1, \cdots, x_N]^T$ is a realization of random vector $\bt X=[X_1, \cdots, X_N]^T$, where $T$ denotes the transposing operation. Matrices are represented by boldface capital characters; whether they contain  random or specific values will be clear from the context. We use $\bt 1_{N}$ to denote the column vector whose $N$ elements are 1; similarly, $\bt 1_{N \times M}$ denotes the all-ones matrix of size $N \times M$. Furthermore, $< . >$ represents the scalar product operation and $\otimes$ the Kronecker product.

\begin{table*}
\begin{center}
\caption{Summary of notation}
\label{tab:notation}
\begin{tabular}{p{1.5cm}l}
  \textbf{Symbol} & \multicolumn{1}{c}{\textbf{Meaning}}   \\
  \toprule
  $\nusers$ & Number of users in the population, denoted by $i \in \{1,\cdots,\nusers\}$\\

  $\thre$ & Threshold of the threshold/pool mix \\
  $\alpha$ & Firing probability of the binomial pool mix \\
  $\sendfreq{i}$ & Probability that a message arriving to the mix comes from user $i$ \\
  $\prob{j}{i}$ & Probability that user $i$ sends a message to user $j$ \\
  $\sendprof{i}$ & Sender profile of user $i$, $\sendprof{i}\doteq[\prob{1}{i}, \prob{2}{i}, \cdots \prob{\nusers}{i}]^T$ \\
  $\recvprof{j}$ & Unnormalized receiver profile of user $j$, $\recvprof{j}\doteq[\prob{j}{1}, \prob{j}{2}, \cdots \prob{j}{\nusers}]^T$ \\
  $\probsall$ & Vector of transition probabilities, $\probsall\doteq[\recvprof{1}^T, \recvprof{2}^T, \cdots, \recvprof{\nusers}^T]^T$\\
  $\uniformi{i}$ & $1-\sum_{j=1}^{\nusers} \prob{j}{i}^2$ \\
  $\meanuniformi$ & $\sum_{i=1}^{\nusers} \sendfreq{i} \uniformi{i}$ \\
  \midrule
  $\rho$ & Number of rounds observed by the adversary \\
  $\send{i}{r}$ ($\recv{j}{r}$)&  Number of messages that the $i$th ($j$th) user sends (receives) in round $r$\\
  $\sendv{r}$ ($\recvv{}{r}$) & Column vector containing elements $\send{i}{r}$ ($\recv{j}{r}$), $i,j=1, \cdots, \nusers$\\
  $\recvv{j}{}$ & Column vector containing elements $\recv{j}{r}$, $r=1, \cdots, \rho$\\
  $\sendm$ & $\rho\times\nusers$ matrix containing all the input observations $\sendm \doteq[\sendv{1}, \cdots, \sendv{\rho}]^T$\\
  $\Hm$ & $\identity{\nusers} \otimes  \sendm$ \\
  $\recvall$ & $\rho\nusers\times 1$ column vector containing all the output observations $\recvall \doteq[\recvv{1}{}^T, \cdots, \recvv{\nusers}{}^T]^T$\\
  $\probest{j}{i}$ & Adversary's estimation of $\prob{j}{i}$ \\
  $\sendprofest{i}$ & Adversary's estimation of user $i$'s sender profile $\sendprof{i}$ \\
  $\recvprofest{j}$ & Adversary's estimation of user $j$'s unnormalized receiver profile $\recvprof{j}$ \\
  $\probsallest$ & Adversary's estimation of transition probabilities vector $\probsall$\\
 \end{tabular}
\end{center}
\end{table*}

\subsubsection{System model}
We study a system in which a population of $\nusers$ users, designated by an index $i  \in \{1, \cdots, \nusers\}$, exchange messages through a high-latency anonymous communication channel. We consider two types of mixes:
\begin{itemize}
  \item \textbf{Threshold Mix}: This mix gathers $\thre$ messages each round, transforms them cryptographically, and outputs them in a random order, hence hiding the correspondence between incoming and outgoing messages.
  \item \textbf{Binomial Threshold Pool Mix}: This pool mix collects $\thre$ messages per round and alters their appearance to avoid bitwise linkability. However, instead of outputting them immediately, messages are placed in a pool and only leave the mix with probability $\alpha$. Otherwise, they stay and get mixed with messages arriving in subsequent rounds.
\end{itemize}

We model the number of messages that user $i$ sends in round $r$ as the random variable $\sendr{i}{r}$, and denote as $\send{i}{r}$ the actual number of messages user $i$ sends in that round. Similarly, $\recvr{j}{r}$ is the random variable that models the number of messages that user $j$ receives in round $r$, and $\recv{j}{r}$ the actual number of messages user $j$ receives in that round. Let $\sendv{r}$ and $\recvv{}{r}$ denote column vectors that contain as elements the number of messages sent or received by all users in round $r$, i.e., $\sendv{r} \doteq [\send{1}{r},\cdots,\send{\nusers}{r}]^T$, and $\recvv{}{r} \doteq [\recv{1}{r},\cdots, \recv{\nusers}{r}]^T$, respectively. When it is clear from the context, the superscript $r$ is dropped. We also group the messages received by user $j$ up to round $\rho$ in vector $\recvv{j}{} \doteq [\recv{j}{1}, \cdots, \recv{j}{\rho}]^T$. Let $\sendm$ denote the $\rho \times \nusers$ matrix containing all the input observations up to round $\rho$, $\sendm  \doteq [\sendv{1}, \sendv{2}, \cdots, \sendv{\rho}]^T$ and let $\recvall$ denote the $\rho \nusers \times 1$ vector containing all the output observations up to round $\rho$, $\recvall \doteq  [\recv{1}{1}, \cdots, \recv{1}{\rho}, \recv{2}{1}, \cdots, \recv{2}{\rho}, \cdots, \recv{\nusers}{1},  \cdots, \recv{\nusers}{\rho}]^T$. Lastly, we define the matrix $\Hm \doteq \identity{\nusers} \otimes  \sendm$ which we shall use when deriving the LSDA estimator in Sect.~\ref{sec:LSDAthre}.

Users in our population send messages to their recipients according to two parameters:
\begin{itemize}
 \item \textbf{Sender profile}: the sender profile of a user represents her communication preferences, i.e., what fraction of her messages is sent to each receiver. We denote the sender profile of user $i$ by vector $\sendprof{i} \doteq [\prob{1}{i}, \prob{2}{i}, \cdots, \prob{\nusers}{i}]^T$, where $\prob{j}{i}$ models the probability that user $i$ sends a message to user $j$.
 We define $\recvprof{j}$ as the column vector containing the probabilities of those incoming messages to the $j$th user, i.e., $\recvprof{j} \doteq [\prob{j}{1}, \prob{j}{2}, \cdots, \prob{j}{\nusers}]^T$. Let $\probsall$ be the vector containing all the transition probabilities, i.e., $\probsall \doteq [\prob{1}{1}, \cdots, \prob{1}{\nusers}, \prob{2}{1}, \cdots, \prob{2}{\nusers}, \cdots, \prob{\nusers}{1}, \cdots, \prob{\nusers}{\nusers}]^T$. With the previous definitions, this vector can be written as $\probsall=[\recvprof{1}^T, \recvprof{2}^T, \cdots, \recvprof{\nusers}^T]^T$.
 We make no assumptions on the shape of users' profiles (i.e., we impose no restrictions on the number of contacts a user may have, nor on how messages are distributed among them), other than $\sendprof{i} \in \probsimplex$, where $\probsimplex$ is the probability simplex in $\mathbb R^{\nusers}$, i.e., $\probsimplex \doteq \left\{ \mathbf{r} \in \mathbb R^{\nusers} : r_i\geq0, \sum_{i=1}^{\nusers} r_i=1\right\}$. 
 \item \textbf{Sending frequency}: the sending frequencies model how often users participate in the system. We denote the sending frequency of user $i$ by $\sendfreq{i}$, where $\sendfreq{i}$ is the probability that a message arriving to the mix comes from user $i$. We make no assumptions on the values of the sending frequencies other than $0\leq \sendfreq{i}\leq 1$ for $i=1,2,\cdots,\nusers$ and $\sum_{i=1}^{\nusers} \sendfreq{i}=1$.
\end{itemize}

Finally, we define $\uniformi{i} \doteq 1-\sum_{j=1}^{\nusers} \prob{j}{i}^2$ and $\meanuniformi \doteq \sum_{i=1}^{\nusers} \sendfreq{i} \uniformi{i}$. The former, $\uniformi{i}$, represents the uniformity of the distribution of user $i$'s sender profile. It ranges from 0, when user $i$ always sends messages to the same user  (i.e., $\prob{k}{i}=1$ for a certain user $k\in\{1,2,\cdots,\nusers\}$, and $\prob{j}{i}=0$ otherwise), to $\frac{\nusers-1}{\nusers}$, when user $i$ sends messages to all the other users equiprobably (i.e., $\prob{j}{i}=\frac{1}{\nusers}$ for $j=1,2,\cdots,\nusers$). The parameter $\meanuniformi$ represents the average uniformity of all users' sender profiles. These parameters shall come in handy in the performance evaluation in Sect.~\ref{sec:eval}.

\subsubsection{Adversary model}
We consider a global passive adversary that observes the system during $\rho$ rounds. She can observe the identity of the senders and receivers that communicate through the mix. Furthermore, she knows all the parameters of the mix (e.g. $\thre$ and/or $\alpha$). As our objective is to illustrate the impact of disclosure attacks on anonymity, we assume that the cryptographic transformation performed by the mix is perfect and thus the adversary cannot gain any information from studying the content of the messages.

The adversary's goal is to uncover communication patterns from the observed flow of messages. Formally, given the observations $\send{i}{r}$ and $\recv{j}{r}$, for $i,j=1,\cdots,\nusers$, and $r=1,\cdots,\rho$, the adversary's goal is to obtain estimates $\probest{j}{i}$ as close as possible to the probabilities $\prob{j}{i}$, which in turn allow her to recover the users' sender and receiver profiles.

\subsection{Working hypotheses}\label{sec:working_hyp}
For the derivation of the LSDA estimator in the next section, we assume that users' choice of recipients is independent for each input message, i.e., the recipient of each message sent by a user $i$ is chosen randomly according to its sender profile $\sendprof{i}$. We also assume that the sender profiles are static, i.e., they do not change within the same round or between different rounds.

For the analysis in Sect.~\ref{sec:performance}, we further assume that the probability that a given message arriving to the mix comes from a certain user is independent for each incoming message, i.e., the number of messages sent by the users in each round can be modeled as a multinomial distribution whose parameters are the sending frequencies. Also, we consider that the sending frequencies are static, i.e., they do not change within the same rounds or between different rounds.

The implications of these assumptions not holding are discussed in Sect.~\ref{sec:discussion}.

\section{A Least Squares approach to Disclosure Attacks on Threshold Mixes}
\label{sec:LSDAthre}

We aim here at deriving a profiling algorithm to recover the sending behavior of users anonymously communicating through a threshold mix based on a least squares approach, under the assumptions explained in Sect.~\ref{sec:working_hyp}. Even though the mix output random variables are conditioned on the input matrix $\bt U$ (or, equivalently, $\bt H$), for the sake of notational simplicity, in this section we will not write such conditioning explicitly.

Our goal is to estimate the users' profiles given the input and output observations, $\send{i}{r}$ and $\recv{j}{r}$ for $i,j=1,\cdots,\nusers$, and all rounds $r=1,\cdots,\rho$. To derive our estimator, we first note that given the vector of probabilities $\probsallreal$ and the input samples in $\sendm$, the output process $\recvr{j}{r}$ for $j=1,\cdots,\nusers$ can be modeled in each round $r\in\{1,\cdots,\rho\}$ as the sum of $\nusers$ multinomials\toremove{, one for each sender}{}:
\begin{equation} \label{eq:output_dist}
 \{\recvr{1}{r}, \cdots, \recvr{\nusers}{r}\}\sim \sum_{i=1}^{\nusers} \text{Multi}\left( \send{i}{r}, \{ \prob{1}{i}, \cdots, \prob{\nusers}{i}\}\right)\,.
\end{equation}

Recall that these output random variables $\recvr{j}{r}$, $j=1, \cdots, N$, $r=1, \cdots, \rho$ are collected in random vector $\recvallr$, of which we observe one realization $\recvall$.  Given this realization $\recvall$ and the input observations $\sendm$, we want to estimate the probability vector $\probsall$. In order to do so, we look for the vector $\probsall$ that minimizes the Mean Squared Error (MSE) between $\recvall$ and $\recvallr$.
Then, our estimator can be formulated as the following constrained least squares problem,
\begin{equation} \label{eq:LSproblem1}
 \probsallest=\operatornamewithlimits{arg\,min}\limits_{ \sendprof{i} \in \probsimplex,\,  i=1,\cdots,N} \text{E}\left\{||\recvall-\recvallr(\probsall)||^2\right\}
\end{equation}
where we have written $\recvallr(\probsall)$ to stress the fact that the output distribution actually depends on the probability vector $\probsall$ (cf. \eqref{eq:output_dist}).
Notice that the constraints are enforced on each sender profile $\sendprof{i}$ to ensure that they are well-defined. Since there exists a one-to-one correspondence between each element in $\sendprof{i}$ for $i=1,\cdots,\nusers$ and each element in $\probsall$, imposing the restrictions over each $\sendprof{i}$ is equivalent to doing so in $\probsall$.

The estimator in \eqref{eq:LSproblem1} minimizes, on average, the squared error over the possible outputs of the system, but does not necessarily minimize the error in the estimation of $\probsall$. The estimator in \eqref{eq:LSproblem1} is actually biased, but expanding the formulation we can find an unbiased and asymptotically efficient estimator of $\probsall$.
First, let $\bt W (\probsall) \doteq \recvallr (\probsall) -\text{E}\{\recvallr (\probsall)\}$, which is a vector containing zero-mean random variables and whose variance is equal to that of $\recvallr(\probsall)$. Using the definition in \eqref{eq:output_dist}, we can write $\text{E}\{\bt Y_j (\recvprof{j})\}=\sendm \cdot \recvprof{j}$, and therefore,  $\text{E}\{\recvallr (\probsall) \}=\left(\identity{\nusers} \otimes  \sendm\right) \probsallcand=\Hm \cdot \probsallcand$.
Then,
\begin{eqnarray} \label{eq:expansion_step}
 \text{E}\left\{||\recvall-\recvallr (\probsall)||^2\right\} &=& \text{E}\left\{||\recvall-\Hm \probsallcand - \bt W (\probsall)||^2\right\} \nonumber \\
     &=& || \recvall - \Hm \probsallcand ||^2+\text{E}\left\{|| \bt W (\probsall)||^2  \right\}
\end{eqnarray}
where we have used that $\text{E}\left\{<(\recvall -\Hm \probsallcand),\bt W (\probsall)>\right\}=\bt 0$ since $\text{E}\{ \bt W (\probsall) \}=\bt 0$.

Removing the term $\text{E}\left\{|| \bt W (\probsall)||^2  \right\}$ from \eqref{eq:expansion_step} leads to an estimator which is asymptotically {\em efficient} when users' behavior is static, in the sense that $\probsallest$ converges to the true profiles as $\rho \rightarrow \infty$. In that case, \eqref{eq:LSproblem1} becomes
\begin{equation} \label{eq:constrained}
 \probsallest=\operatornamewithlimits{arg\,min}\limits_{ \sendprof{i} \in \probsimplex,\,  i=1,\cdots,N} ||\recvall-\Hm \probsallcand||^2\,.
\end{equation}

\subsection{Unconstrained Least Squares Estimation}
We first propose an unconstrained estimator which is still asymptotically efficient and amenable to an in-depth performance analysis, as we will show in Sect.~\ref{sec:performance}. It is well-known that, for the unconstrained case, the solution of \eqref{eq:constrained} is provided by the Moore-Penrose pseudoinverse \cite{Sch91}:
\begin{equation} \label{eq:formidableinversion}
\probsallest = (\Hm^T \Hm)^{-1} \Hm^T \recvall \,.
\end{equation}

At first sight, it might look that the matrix inversion needed in \eqref{eq:formidableinversion} is formidable: the matrix $\Hm^T \Hm$ has size $\nusers^2 \times \nusers^2$. However, its block-diagonal structure allows for a more affordable solution:
\begin{equation*}
\Hm^T \Hm =  ( \identity{\nusers} \otimes \sendm)^T \cdot \identity{\nusers} \otimes \sendm = \identity{\nusers} \otimes (\sendm^T \sendm)
\end{equation*}
and, hence,
\begin{equation*}
 (\Hm^T \Hm)^{-1} =  \identity{\nusers}  \otimes (\sendm^T \sendm)^{-1}
\end{equation*}
where $\sendm^T \sendm$ of size $\nusers \times \nusers$ is assumed to have full rank.

The decoupling above allows us to write a more efficient solution. The least squares estimate $\recvprofest{j}$ for the $j$th probability vector can be written as
\begin{equation} \label{eq:LSDAthre}
\recvprofest{j} =  (\sendm^T \sendm)^{-1} \sendm^T \recvv{j}{},\ \ j=1, \cdots, \nusers \,.
\end{equation}
Notice that, as a consequence of removing the constraints, the obtained sender profiles are not guaranteed to lie in the probability simplex $\probsimplex$. In fact, some of the estimated probabilities $\probest{j}{i}$ will often be negative, usually corresponding to receivers $j$ that are not contacts of user $i$.

Note that the matrix operations performed by LSDA have much smaller computational requirements than the round-by-round processing carried out by previous attacks \cite{TGPV08,DT09}. This decrease in computation comes at the cost of memory: LSDA has to deal with large matrices. When memory is an issue, the Recursive Least Squares (RLS) algorithm~\cite{haykin}, that computes the least squares solution by processing the observed rounds recursively, can reduce the requirements of the attack considerably.


\subsubsection{The Statistical Disclosure Attack as an LS estimator}
We now show that the original Statistical Disclosure Attack~\cite{Danezis03} in fact corresponds to a particular case of the proposed LSDA estimator. Here, the first user (Alice) is supposed to send only one message to an unknown recipient chosen uniformly from a set of $\nfriends$ contacts. The other users are assumed to send messages to recipients chosen uniformly from the set of all users. The target is to determine the set of contacts of Alice.

From these considerations, for a given round $r$ where Alice does send a message, we have that $\send{1}{r}=1$ and $\sum_{i=2}^{\nusers} \send{i}{r} = (\thre-1)$, and all the transition probabilities $\prob{j}{i}$, for $i \geq 2$, $j=1, \cdots, \nusers$, are known to be equal to $1/\nusers$. If we suppose that in all rounds Alice transmits a message, we will have a vector $\recvall$ which contains the $\rho \cdot \nusers$ observations, $\sendprof{1}$ is unknown, and all $\sendprof{i}$, $i=2, \cdots, \nusers$ are known. The unconstrained LSDA estimator can be broken down into subproblems in which we seek $\recvprof{j}$, for all $j=1, \cdots, \nusers$,  such that
\begin{equation}
\label{eq:subproblems}
||\recvv{j}{} - \sendm \recvprof{j}||^2
\end{equation}
is minimized. Noticing that for each $\recvprof{j}$ only $\prob{j}{i}$ is unknown, we can write the equivalent problem of finding $\prob{j}{1}$ such that
\begin{equation}
||\recvv{j}{} - \sendm' \bt p'_j - \prob{j}{1} \sendm_1||^2
\end{equation}
is minimized, where $\sendm'$ is obtained from $\sendm$ by deleting its first column, itself denoted by $\sendm_1$, and where $\bt p'_j$  is obtained from $\recvprof{j}$ after deleting its first element.

Then, the LS solution is
\begin{equation}
\probest{j}{1} = (\sendm_1^T \sendm_1)^{-1} \sendm_1^T (\recvv{j}{} - \sendm' \bt p'_j)\,.
\end{equation}

From the fact that $\sendm_1=\bt 1_\rho$ (as Alice sends one and only one message per round), it follows that $\sendm_1^T \sendm_1 = \rho$. On the other hand, all elements in $\bt p'_j$ take the value $1/\nusers$ and the matrix $\sendm'$ is such that the sum of the elements in each row is $(\thre-1)$; therefore,
\begin{equation} \label{eq:SDA}
\probest{j}{1} = \frac{1}{\rho} \sum_{r=1}^\rho \recv{j}{r} - \frac{(\thre-1)}{\nusers}, \ \ j=1, \cdots, \nusers
\end{equation}
which coincides with Danezis' SDA estimate~\cite{Danezis03}.

The LSDA estimator differs from the original SDA estimator in that it does not make any underlying assumption on the transition probabilities and that it simultaneously solves for the entire matrix of transition probabilities.

\subsubsection{Performance analysis with respect to the system parameters}
\label{sec:performance}

Next, we assess the performance of our unconstrained solution in \eqref{eq:LSDAthre} for the working hypothesis explained in Sect.~\ref{sec:working_hyp}. This will serve to understand the influence of the system parameters on the knowledge that an adversary can gain by applying our algorithm. 
To the best of our knowledge, this is the first in-depth analysis of how the system parameters affect the performance of an attack on mixes. We aim at deriving a theoretical expression for the MSE in the estimation of the sender profile of user $i$, which we define as
\begin{equation} \label{eq:MSEi}
  \MSEi\doteq\text{E}\{ || \sendprof{i}-\sendprofest{i}||^2\}=\sum_{j=1}^{\nusers} \text{E}\left\{ \left(\prob{j}{i}-\probest{j}{i}\right)^2 \right\}\,.	
\end{equation}
 We carry out this derivation in Appendix~\ref{sec:appendix_thre}, obtaining
\begin{equation} \label{eq:MSEthre}
 \MSEi\approx \frac{1}{\rho} \left\{ \left(\sendfreq{i}^{-1}-1\right)\left(1-\frac{1}{\thre}\right)\meanuniformi + \frac{\sendfreq{i}^{-1}}{\thre} \cdot \uniformi{i}\right\}\,.
\end{equation}

It is useful to interpret \eqref{eq:MSEthre} in terms of the users' sending frequency $\sendfreq{i}$ and the profile uniformity $\uniformi{i}$. First, the error of LSDA increases with $\sendfreq{i}^{-1}$, since it becomes harder to estimate the behavior of a user when that user rarely participates in the system. The $\MSEi$ also increases with $\uniformi{i}$ and $\meanuniformi$. This also makes sense, since given a certain amount of observations it is harder to infer the sending profiles when the behavior of the users in the system is highly random (large $\uniformi{i}$ or $\meanuniformi$) than when the users are more predictable (low $\uniformi{i}$ or $\meanuniformi$).
The MSE also decreases as $1/\rho$ with the number of rounds $\rho$; this implies that the unconstrained LS estimator is asymptotically efficient as $\rho \rightarrow \infty$. Even though this is somewhat to be expected, notice that other estimators might not share this desirable property, as we will experimentally confirm in Sect.~\ref{sec:eval}.

Note that our expression for the theoretical MSE of LSDA in \eqref{eq:MSEthre} is an approximation due to the simplification \eqref{eq:LLN_Rx} made during its derivation in Appendix~\ref{sec:appendix_thre}. Therefore, there will be a discrepancy between our MSE formula and the real MSE of the LSDA estimator which increases with the number of users in the system $\nusers$ and decreases with $\rho$.

To better understand the performance of the estimator, we now derive a rough approximation of \eqref{eq:MSEthre}. Normally, when the number of users is large, we can expect the sender frequencies of the users to be low, i.e., $\sendfreq{i}^{-1}\gg1$. Also, if we assume that users have many contacts without any specific preference to any of them, then $\sum_{j=1}^{\nusers}\prob{j}{i}^2 \ll 1$ or, equivalently, $\uniformi{i}\approx 1$. In this case, \eqref{eq:MSEthre} can be approximated as
\begin{equation} \label{eq:MSEthre_approx}
 \MSEi \approx \frac{\sendfreq{i}^{-1}}{\rho}\,.
\end{equation}
Although this is a rough approximation, it shows the dominant parameters that affect the performance of the attack. Furthermore, it shall be useful when comparing the performance of LSDA for threshold mixes with that of pool mixes.

\subsection{Constrained Least Squares Estimation}
\label{sec:clsda}

We now derive an estimator for the constrained problem in \eqref{eq:constrained}. Note that one could reduce the error of the unconstrained estimator \eqref{eq:LSDAthre} by just setting the negative probabilities to zero. The $\sum_j{\prob{j}{i}}=1$ constraint could be later ensured by normalizing the profile, but this normalization has to be performed without information and hence the estimation is not guaranteed to be optimal.

We recall that the constraints are $0 \leq \prob{j}{i} \leq 1,$ for all $ i,j=1, \cdots, \nusers$, and  $\sum_{j=1}^{\nusers} \prob{j}{i}=1,$ for all  $i=1, \cdots, \nusers$. One might think of imposing such constraints to the decoupled optimization problems \eqref{eq:subproblems} for each $j$. Unfortunately, while the optimization is performed with respect to $\recvprof{j}$, each of the previous sum constraints is given in terms of $\sendprof{i}$. Hence, if those constraints are to be enforced, then the optimization problems can no longer be decoupled.

An alternative solution consists in solving the problem in an iterative fashion. If we define $\transprobest=[\sendprofest{1}, \sendprofest{2}, \cdots, \sendprofest{\nusers}]$ and $\bt V \doteq [\recvv{1}{}, \recvv{2}{}, \cdots, \recvv{\nusers}{}]$, then the {\em unconstrained} solution in \eqref{eq:LSDAthre} can be written in a more compact form as
\begin{equation} \label{eq:compact_problem}
 \transprobest^T=(\sendm{}^T\sendm)^{-1}\sendm^T \bt V\,.
\end{equation}
Using a gradient descent algorithm, it is possible to solve \eqref{eq:compact_problem} by iteratively updating matrix $\transprobest$ as
\begin{equation} \label{eq:landweber}
 \transprobest_{new}=\transprobest_{old}-\stepsize \cdot \sendm^T \left( \sendm \cdot \transprobest_{old} - \bt V\right)\,.
\end{equation}
Here, the stepsize $\stepsize$ is chosen such that $0<\stepsize<2/\lambda_{max}$, where $\lambda_{max}$ is the largest eigenvalue of matrix $\sendm^T \sendm$. The {\em constrained} solution can be obtained by projecting each column of $\transprobest$ onto the probability simplex $\probsimplex$ after each iteration. This solution, which we will denote C-LSDA, outperforms the unconstrained one as we will show in Sect.~\ref{sec:eval}.

As a final remark, the constraints make a performance analysis similar to that in Section~\ref{sec:performance} much more cumbersome. The analysis of such solution is left as subject for future research, but we note that the MSE for the unconstrained version approximated in \eqref{eq:MSEthre} constitutes an upper bound on the MSE of the constrained variant. 
\section{A Least Squares approach to Disclosure Attacks on Pool Mixes}
\label{sec:LSDApool}

In this section, we show how to extend the Least Squares Disclosure Attack and the analysis of its performance to the threshold binomial pool mix, which was described in Sect.~\ref{sec:sysmodel}, working under our hypotheses in Sect.~\ref{sec:working_hyp}. We note, however, that the principles behind the attack make it easily adaptable to other mixing strategies (e.g., timed mixes), as shown in~\cite{WIFS12}. The main difference of the pool mix with respect to the threshold mix arises from the fact that some messages stay in the pool so it is no longer possible for the adversary to know how many messages from user $i$ leave the mix in round $r$.

\begin{figure}[!t]
\centering
  \includegraphics[width=2.2in]{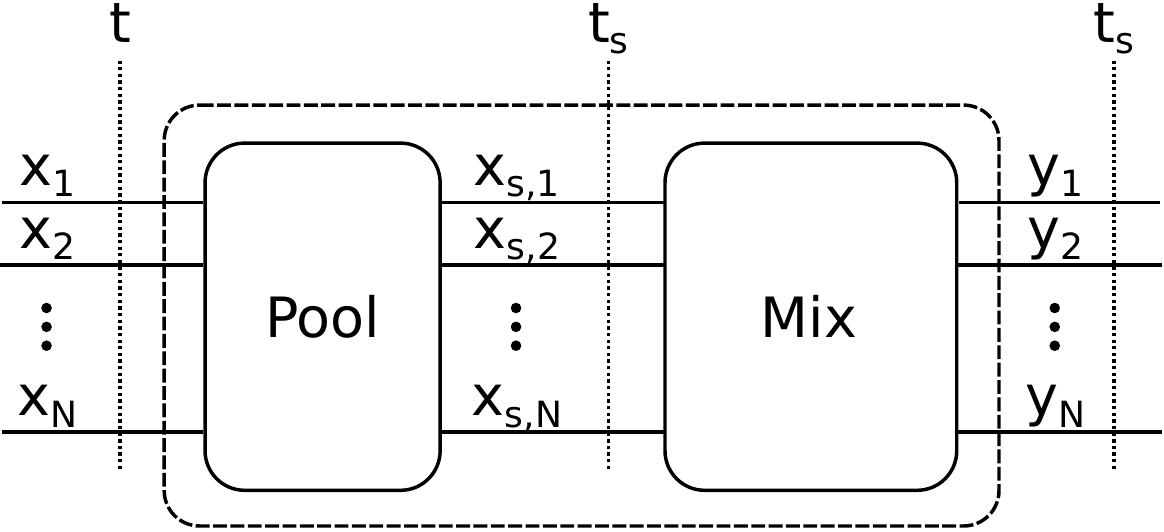}
\caption{Abstract model of the pool mix, represented as a combination of two blocks: the pool block, which delays the messages, and the mix block, which operates as a standard threshold mix.}
\label{fig:pool_mix}
\end{figure}

In order to make the derivation of the estimator easier, we abstract the threshold binomial pool mix as a combination of a pool block followed by a mixing block, as depicted in Fig.~\ref{fig:pool_mix}. The pool block stores messages until it has received $\thre$ of them, and then outputs each with probability $\alpha$, leaving the remaining for subsequent rounds. The messages that leave the pool block traverse the mix block, which changes their appearance and forwards them to their receivers. Note that the pool mix always receives $\thre$ messages each round, while the number of messages that leave, denoted by $t_s$ in the figure, is variable. To distinguish between the number of messages from user $i$ that enter and leave the pool block in round $r$, we will respectively use $\send{i}{r}$ and $\sendpool{i}{r}$, as shown in Fig.~\ref{fig:pool_mix}. The adversary is only able to observe vectors $\sendv{r}$ and $\recvv{}{r}$, while the number of messages from each sender that leave the mix in each round, modeled by the random vector $\sendpoolvr{r}$, is unknown to the attacker. We let $\sendpoolmr^T \doteq [\sendpoolvr{1}, \sendpoolvr{2}, \cdots, \sendpoolvr{\rho}]$.

We assume that at the time the adversary starts her observation the pool contains $m$ messages whose sender is unknown. Then, the messages in $\sendpoolr{i}{r}$ may come from two sources: the initial $m$ messages in the pool and the messages sent by user $i$ in the current or earlier rounds. We will use $\sendpoolnoise{i}{r}$ to model the number of messages from user $i$ that were initially in the pool and leave the mix in round $r$, and $\sendpoolr{i}{r,k}$ to model the number of messages sent by user $i$ in round $k$ that leave the mix in round $r$. Thus, the total number $\sendpoolr{i}{r}$ of messages from user $i$ that leave the mix in round $r$ can be written as
\begin{equation} \label{eq:pool_contributions}
 \sendpoolr{i}{r}=\sum_{k=1}^{r} \sendpoolr{i}{r,k}+\sendpoolnoise{i}{r}\,
\end{equation}
where each of the contributions $\sendpoolr{i}{r,k}$ for $k=1,\cdots,r$ and $\sendpoolnoise{i}{r}$ are independent. The random variables $\{\sendpoolr{i}{k,k}, \sendpoolr{i}{k+1,k},\cdots, \sendpoolr{i}{k+l,k}, \cdots\}$ can be modeled together as a multinomial distribution with $\send{i}{k}$ trials and probabilities $\{\alpha, \alpha(1-\alpha), \cdots, \alpha(1-\alpha)^l, \cdots\}$.

We derive the LSDA estimator for the pool mix problem as we did in Sect.~\ref{sec:LSDAthre} for the threshold mix. Again, for notational simplicity, we will not write the conditioning of $\sendpoolmr$ and $\recvallr$ on $\sendm$. Given the vector of probabilities $\probsallreal$ and the messages that leave the pool in each round $\sendpoolm$, we can model the output process as
\begin{equation} \label{eq:output_dist_pool}
 \{\recvr{1}{r}, \cdots, \recvr{\nusers}{r}\}\sim \sum_{i=1}^{\nusers} \text{Multi}\left( \sendpool{i}{r}, \{ \prob{1}{i}, \cdots, \prob{\nusers}{i}\}\right)\,.
\end{equation}
In this case, however, the values in $\sendpool{i}{r}$ are not observable. To derive our estimator following the approach in Sect.~\ref{sec:LSDAthre}, we need to compute the expected value of $\recvallr(\probsall)$. In this case,
\begin{equation} \label{eq:expZpool}
 \text{E}\{ \bt Y_j(\recvprof{j}) \}
    =\text{E}\{ \text{E}\{ \bt Y_j(\recvprof{j}) | \sendpoolm \} \}
    =\text{E}\{ \sendpoolmr \}\cdot \recvprof{j} =\sendpoolmest \cdot \recvprof{j}
\end{equation}
where each element of $\sendpoolmest\doteq\text{E}\{ \sendpoolmr\}$ can be computed, using~\eqref{eq:pool_contributions}, as
\begin{eqnarray}
\sendpoolest{i}{r} &=& \text{E}\{ \sendpoolr{i}{r} \} =  \sum_{k=1}^{r} \text{E}\{ \sendpoolr{i}{r,k} \} + \text{E}\{ \sendpoolnoise{i}{r} \} \nonumber \\
&=&  \sum_{k=1}^{r} \send{i}{k} \alpha (1-\alpha)^{r-k}  + m \sendfreqest{i} \alpha (1-\alpha)^{r-1}\,.
\label{eq:abroad}
\end{eqnarray}
Here, $\sendfreqest{i}$ represents the adversary's estimation of the probability that each of the $m$ messages in the initial pool corresponds to user $i$. An attacker can estimate those values by observing the system for a while. Anyhow, the influence of the initial $m$ messages in the estimation of the profiles diminishes quickly as the number of rounds observed increases. For implementation purposes, a more convenient way of writing \eqref{eq:abroad} is the following recursive equation
\begin{equation}
\sendpoolest{i}{r+1}=(1-\alpha) \sendpoolest{i}{r} + \alpha \send{i}{r+1},\ \ r=1, \cdots, \nusers
\end{equation}
where $\sendpoolest{i}{1}$ is initialized to $\send{i}{1}+m \sendfreqest{i}$.

For compactness, we will find it useful to define the following {\em convolution matrix}
\begin{equation}
\B \doteq \left[ \begin{array}{cccccc}
\alpha & 0 &  0 & \cdots & 0\\
\alpha (1-\alpha) & \alpha & 0 & \cdots & 0\\
\alpha (1-\alpha)^2 & \alpha (1-\alpha) & \alpha & \cdots & 0\\
\vdots & \vdots & \vdots & \cdots & \vdots \\
\alpha (1-\alpha)^{\rho-1} & \alpha (1-\alpha)^{\rho-2} &  \alpha (1-\alpha)^{\rho-3}& \cdots & \alpha\\
\end{array} \right]
\end{equation}
Then, we can write
\begin{equation} \label{eq:Us}
\sendpoolmest = \B \cdot \left( \sendm + \Nzero \right)
\end{equation}
where the matrix $\Nzero$, which accounts for the average initial state of the mix, is such that the $i$-th entry in the first row takes the value $m \sendfreqest{i}$, while all the remaining elements are zero.

Our LSDA estimator in the pool mix, following the derivations given in the threshold mix scenario~\eqref{eq:constrained}, can be formulated as the following constrained problem:
\begin{equation} \label{eq:constrained_pool}
 \probsallest=\operatornamewithlimits{arg\,min}\limits_{ \sendprof{i} \in \probsimplex,\,  i=1,\cdots,N} ||\recvall-\Hmpoolest \probsallcand||^2
\end{equation}
where $\Hmpoolest\doteq\identity{\nusers} \otimes \sendpoolmest$.

\subsection{Unconstrained Least Squares Estimation on Pool Mixes}

The solution for the unconstrained case is given by
\begin{equation} \label{eq:LSDApool}
\recvprofest{j} =  (\sendpoolmest^T \sendpoolmest )^{-1} \sendpoolmest^T  \recvv{j}{},\ \ j=1, \cdots, \nusers
\end{equation}
where $\sendpoolmest=\B \cdot \left( \sendm + \Nzero \right)$. Notice that for the standard threshold mix, which corresponds to $\alpha=1$, $m=0$, we have that $\B = \identity{\rho}$, $\Nzero = \bt 0$, so $\sendpoolmest = \sendm$, and both solutions coincide.

\subsubsection{Performance analysis with respect to the system parameters}
The performance analysis of the LSDA estimator in the pool mix for static user behavior is carried out in Appendix~\ref{sec:appendix_pool}, where it is shown that
\begin{equation} \label{eq:MSEpool}
\begin{array}{lcl}
\MSEi &\approx& \displaystyle\frac{1}{\rho}\left\{ (\sendfreq{i}^{-1}-1)\left[ \meanuniformi \left(\frac{1}{\alpha_r}-\frac{1}{\thre}\right) + \left(\frac{1}{\alphaq} - \frac{1}{\alpha_r}\right)\right]\right. \vspace{0.2cm}\\
&+&\displaystyle\left.\frac{\sendfreq{i}^{-1}}{\thre}\cdot \uniformi{i} \right\}
\end{array}
\end{equation}
where $\alpha_q\doteq \displaystyle\frac{\alpha}{2-\alpha}$ and $\alpha_r \doteq \displaystyle\frac{\alpha(2-\alpha)}{2-\alpha(2-\alpha)}$. This approximation is asymptotically tight as $\rho \rightarrow \infty$. Moreover, when $\alpha=1$ we recover \eqref{eq:MSEthre}. 

Comparing the MSE estimator in \eqref{eq:MSEpool} with the threshold mix one \eqref{eq:MSEthre}, we can conclude that the difficulty of learning the profiles is always larger in the pool mix, since $1/\alpha_r \geq 1$ and $\displaystyle\left(1/\alpha_q - 1/\alpha_r \right)\geq 0$. In the particular case that we have analyzed for the threshold mix, where we have assumed that $\sendfreq{i}^{-1}\gg1$ and $\uniformi{i} \approx 1$, \eqref{eq:MSEpool} can be approximated by
\begin{equation}
 \MSEi \approx \frac{\sendfreq{i}^{-1}}{\rho} \cdot \frac{2-\alpha}{\alpha}\,.
\end{equation}
Comparing this approximation with \eqref{eq:MSEthre_approx} we can conclude that the pool mix requires {\em approximately} $(2-\alpha)/\alpha$ times more rounds to achieve the same MSE. Of course, this comes at the price of an increased delay; in fact, it can be shown that the {\em average} delay for a message introduced by the pool, measured in rounds, is $(1-\alpha)/\alpha$.

\subsection{Constrained Least Squares Estimation on Pool Mixes}

We remark that it is also possible to implement a constrained version of the estimator in the pool mix which forces the \toremove{estimated}{} profiles to lie in the feasible set ${\mathcal P}$ by just replacing matrices $\sendm$ by $\sendpoolmest$ in \eqref{eq:landweber}. As we will confirm in the evaluation section, the constrained version outperforms the unconstrained one.

\section{Evaluation}
\label{sec:eval}

\subsection{Experimental setup}
\label{sec:exp_setup}

We evaluate the effectiveness of LSDA against synthetic anonymized traces created by a simulator written in the Matlab language.\footnote{The code will be made available upon request.} We simulate a population of $\nusers$ users with $\nfriends$ contacts each to whom they send messages following a Zipf distribution. We have chosen the Zipf distribution as a particular case of the power law probability distribution often used to model social networks~\cite{PowerLaw}. We note that both the LSDA estimator and the theoretical approximation of its performance~\eqref{eq:MSEthre} are not restricted to any specific shape of sender profiles and therefore they would be applicable in any other case (e.g.,~\cite{PT12}). In our baseline experiment, users send messages with the same frequency ($\sendfreq{i}=1/N$ for all $i$) although we also simulate the case where users do not participate evenly in the system. For all the experiments in this section, we keep the sender profiles and sending frequencies constant between rounds, as assumed by our model in Sect.~\ref{sec:working_hyp}.

In the first part of the evaluation, messages are anonymized using a threshold mix with threshold $\thre$ and in the second part using a binomial pool mix where in each round $\thre$ messages arrive to the mix and each message in the pool has a probability $\alpha$ of leaving the mix. We evaluate the results for the case that the adversary observes $\rho$ rounds of mixing.
Table~\ref{tab:parameters} summarizes the values of the parameters used in our experiments, where bold numbers indicate the parameters of the baseline experiment.

\begin{table}
\begin{center}
\caption{System parameters used in the experiments.}
\label{tab:parameters}
\begin{tabular}{l c}
   \textbf{Param} & \multicolumn{1}{c}{\textbf{Value}}   \\
					\cline{1-2}
 $\rho$			& $\{{\bf 10\,000},20\,000,\ldots,100\,000\}$ \\
 $\nusers$ 		& $\{25, 50, 75, {\bf 100},200,300,\ldots,1\,000\}$ \\
 $\nfriends$ 		& $\{10,20,{\bf 25},30,40,50,\cdots,100\}$ \\
 $\sendfreq{i}$ 	& $\{{\bf uniform} (\sendfreq{i}=1/\nusers), Zipf (\sendfreq{i}=i^{-1}/\sum_{k=1}^{\nusers}\sendfreq{k})\}$ \\
				      \cline{1-2}
 $\thre$		& $\{1,2,5,{\bf 10},20,30,40,50\}$ \\
 $\alpha$ 		& $\{0.1,0.2,0.3,0.4,0.5,0.6,0.7,0.8,0.9,1\}$ \\
 \end{tabular}
\end{center}
\end{table}

We compare the effectiveness of the unconstrained (LSDA) and constrained (C-LSDA) versions of our attack when profiling users with respect to the following attacks:
\begin{itemize}
 \item SDA-MD: we have implemented Mathewson and Dingledine's version of the Statistical Disclosure Attack as explained in~\cite{MD04} for the threshold mix. However, we have found that the extension of the attack to the pool mix in~\cite{MD04} returned incorrect profile estimations due to scaling issues in its formulation. We have corrected the formula so it does estimate the profiles properly while keeping the philosophy of the attack (we have used the formula for $\mathtt{SDA1}$ in~\cite{GlobalSIP} using matrix $\sendpoolmest$ instead of $\sendm$).

 \item PMDA: the Perfect Matching Disclosure Attack, proposed by Troncoso et al.~in~\cite{TGPV08}, estimates the users profiles in two steps: first, starting with an initial estimation of the profiles (e.g., SDA), it computes the most probable correspondence between input and output messages in each round. Then, using that information, it builds a new estimation of the profiles based on a weight parameter $z$, which is $z=0.5$ in~\cite{TGPV08}. We have implemented the attack using $z=1$, since we have observed that this yields the best results. Note that, because of this, the results of PMDA in this paper differ from those in~\cite{PT12}.

 \item Vida: the Bayesian inference attack proposed by Danezis and Troncoso in~\cite{DT09} allows to draw samples from the distribution of the sender profiles given the observations. We have fixed errors in the implementation of the algorithm in~\cite{DT09,PT12}. Also, in order to compare with the rest of the attacks, we have decided to compute the average of the samples obtained using the Bayesian inference attack, since the MSE of this average profile is always lower than the average MSE of the individual samples.
\end{itemize}

\subsection{Performance metrics}
We recall that the goal of the adversary is to estimate the values $\prob{j}{i}$ with as much accuracy as possible. We define two metrics to illustrate the profiling accuracy of the attacks. The \emph{Mean Squared Error per sender profile} ($\MSEi$), previously defined in~\eqref{eq:MSEi}, measures the squared error between the estimated sender profile of user $i$, $\sendprofest{i}$, and the real sender profile $\sendprof{i}$.
Secondly, the \emph{average Mean Squared Error per transition probability} measures the average squared error in each transition probability $\probest{j}{i}$,
\begin{equation} \label{eq:MSEp}
    \MSEp\doteq \frac{1}{\nusers^2} \sum_{i=1}^{\nusers} \MSEi = \frac{1}{\nusers^2} \sum_{i=1}^{\nusers} \sum_{j=1}^{\nusers} \text{E}\left\{ \left(\prob{j}{i}-\probest{j}{i}\right)^2\right\}\,.
\end{equation}
Both MSEs measure the amount by which the values output by the attack differ from the actual value to be estimated. The smaller the MSE, the better is the adversary's estimation of the users' actual profiles.

For each of the studied set of parameters ($\rho$, $\nusers$, $\uniformi{i}$, $\sendfreq{i}$, $\thre$, $\alpha$) we record the sets of senders and receivers during $\rho$ rounds and compute the $\MSEp$ (or $\MSEi$) for each of the attacks. We repeat this process 100 times and plot the average of the results in our figures.

\subsection{Results: Threshold mix}

We first study the effectiveness of LSDA in profiling messages anonymized using a threshold mix in different scenarios.

\subsubsection{Performance with respect to the number of rounds $\rho$}

\begin{figure}[!t]
\centering
  \includegraphics[width=2.8in]{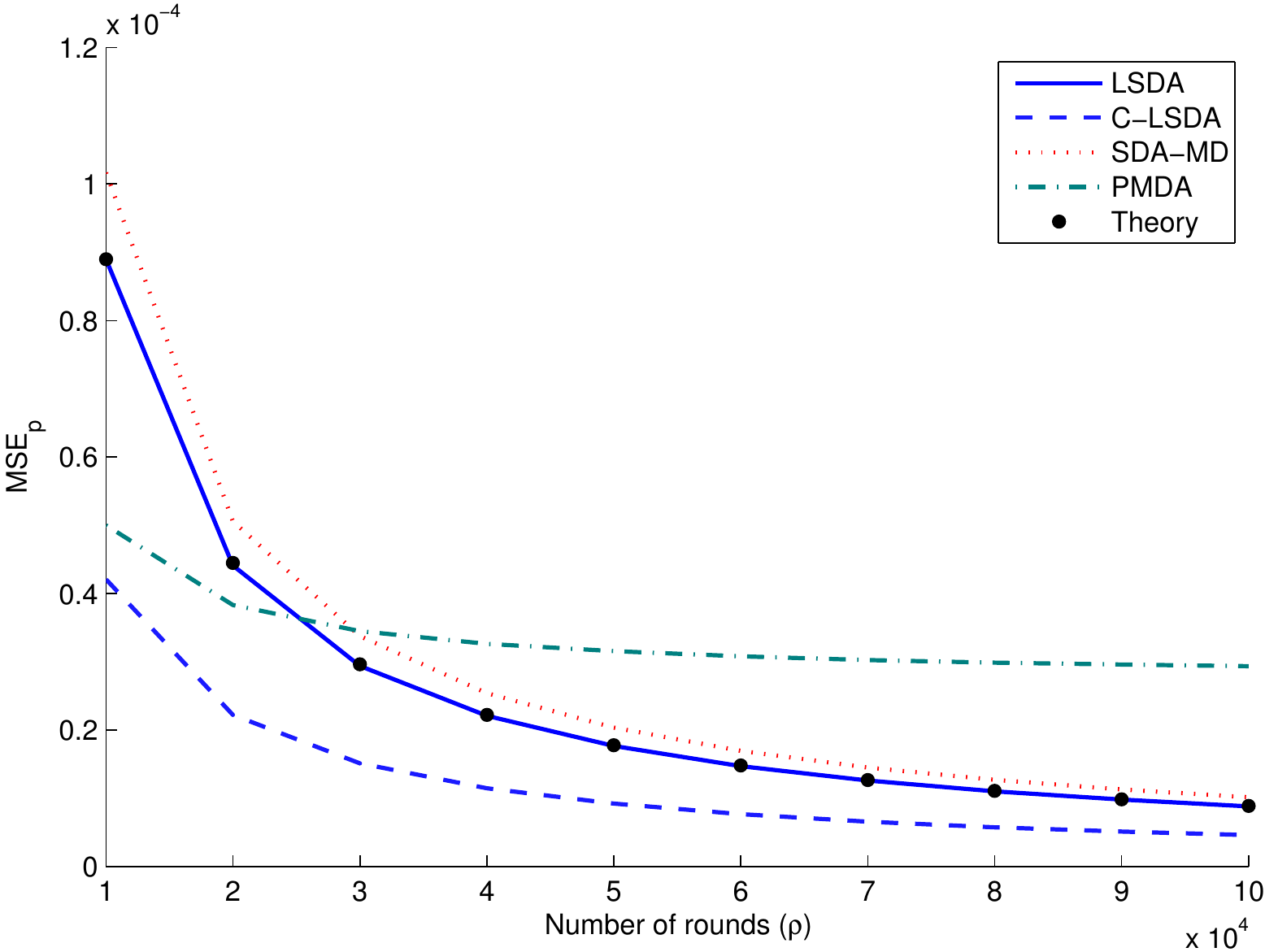}
\caption{$\MSEp$ evolution with the number of rounds in the system $\rho$ ($\nusers=100$, $\nfriends=25$, $\sendfreq{i}=1/\nusers$, $\thre=10$).}
\label{fig:with_rho_thre}
\end{figure}

As we discuss in \ref{sec:performance}, the number of observed rounds $\rho$ has a dominant role in the estimation error incurred by LSDA. We plot in Fig.~\ref{fig:with_rho_thre} the MSE per transition probability $\MSEp$ for LSDA, C-LSDA, SDA-MD and PMDA.

The constrained LSDA obtains the best results, although the unconstrained variant and Mathewson and Dingledine's version of SDA follow up closely. These three estimators are unbiased and asymptotically efficient, i.e., $\MSEp\rightarrow 0$ when $\rho \rightarrow \infty$. Furthermore, we can see how the approximation in \eqref{eq:MSEthre}, represented by $\bullet$ in the figure, reliably describes the decrease in the profile estimation error as more information is made available to the adversary.
On the other hand, PMDA relies on an initial estimation of the profiles which, for $\nusers=100$, is far from reality. Therefore, even when the number of rounds observed increases, PMDA is not able to improve as effectively as the other attacks.

\subsubsection{Performance with respect to the number of users $\nusers$}

\begin{figure*}[!t]
  \centering
  \subfloat[$\rho=10000$]{\includegraphics[width=2.8in]{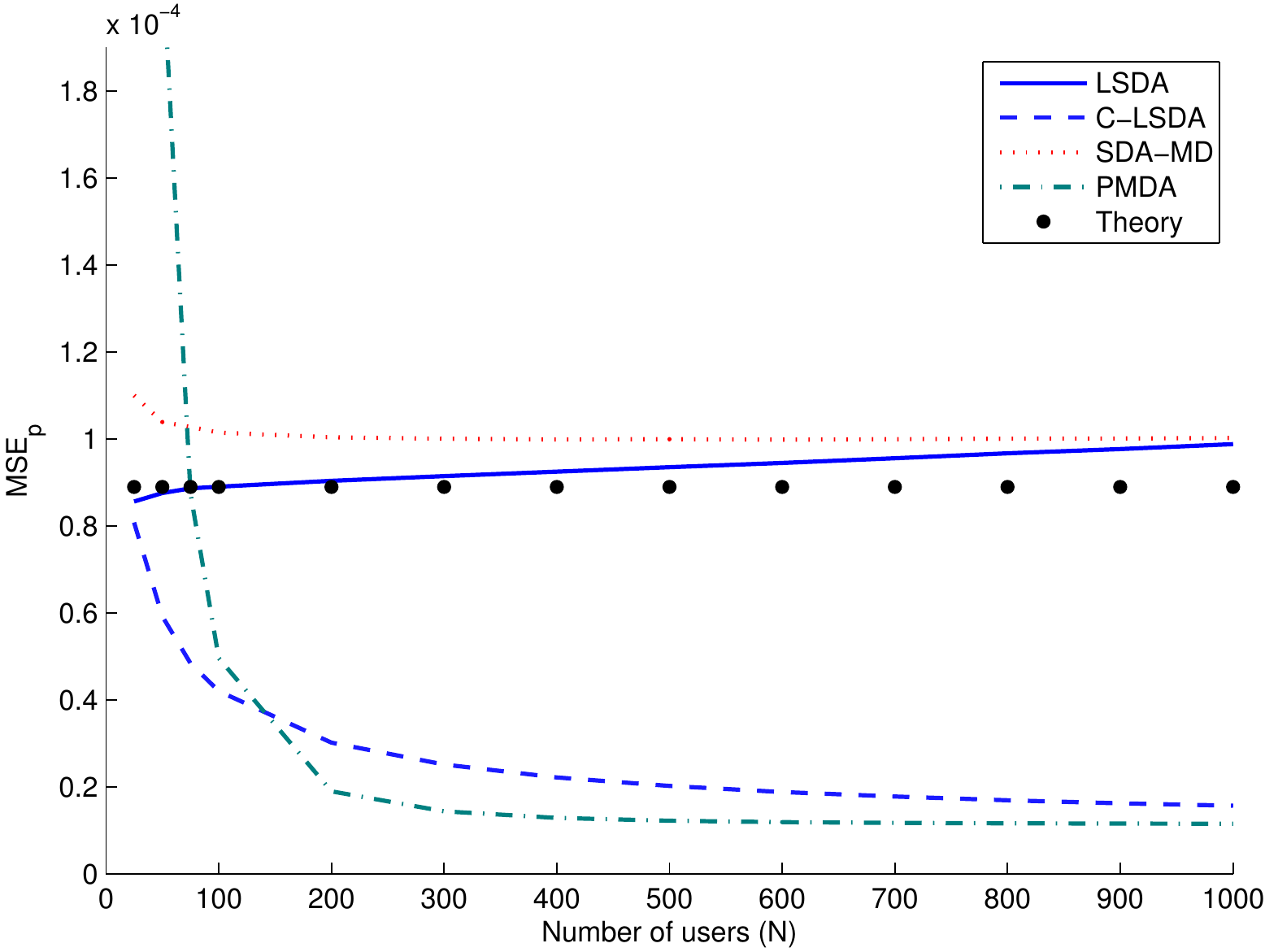} \label{fig:with_N_thre1}}
  \hfil
  \subfloat[$\rho=100000$]{\includegraphics[width=2.8in]{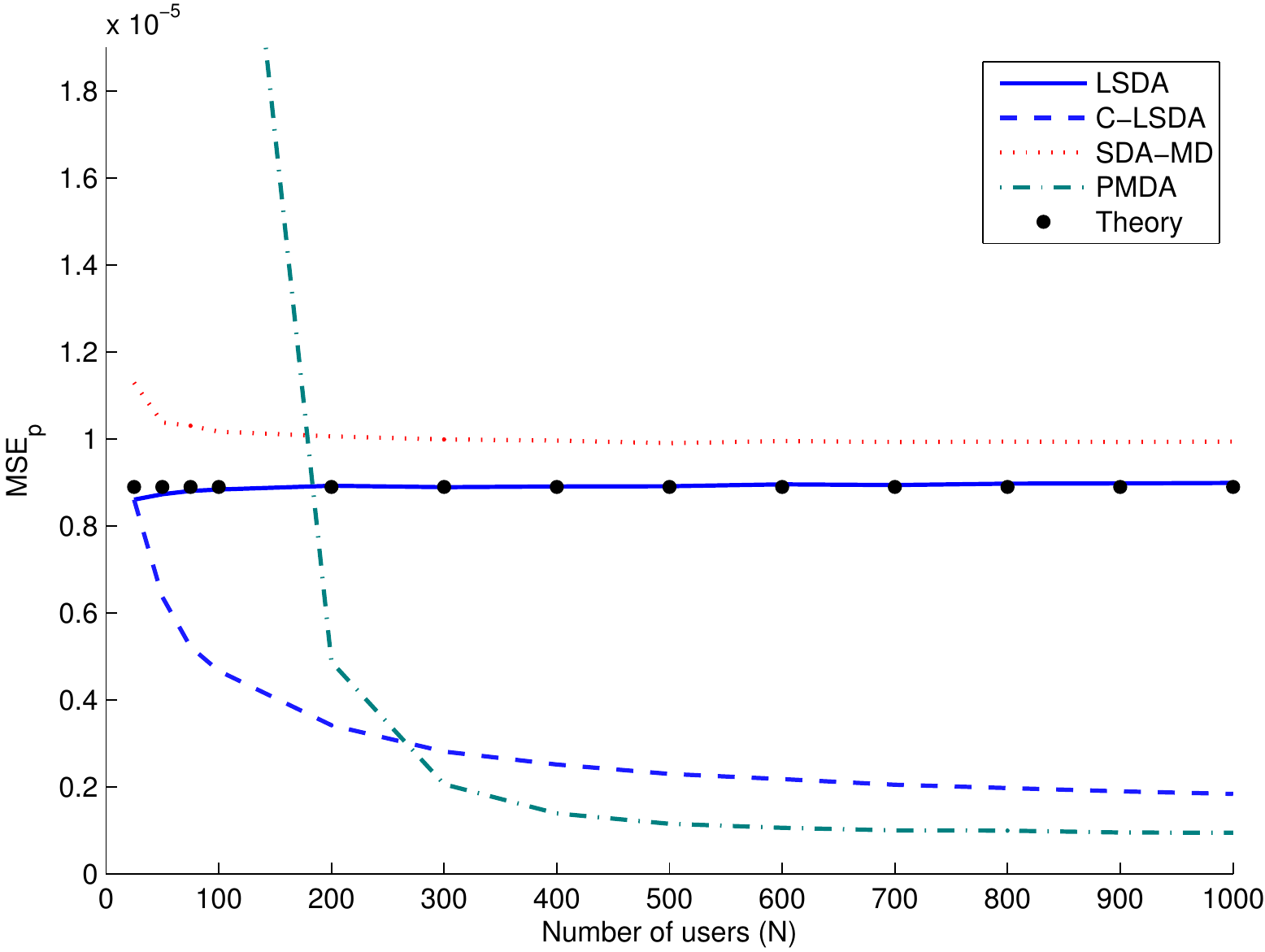} \label{fig:with_N_thre2}}
  \caption{$\MSEp$ evolution with the number of users in the system $\nusers$ ($\rho=10\,000, 100\,000$,  $\nfriends=25$, $\sendfreq{i}=1/\nusers$, $\thre=10$).}
  \label{fig:with_N_thre}
\end{figure*}

Next, we study the influence of the number of users in the system on the estimation error. The results are shown in Fig.~\ref{fig:with_N_thre1} for $\rho=10\,000$ and Fig.~\ref{fig:with_N_thre2} for $\rho=100\,000$. Note that we have adjusted the vertical axis of Fig.~\ref{fig:with_N_thre2} so that it is 10 times smaller than that in Fig.~\ref{fig:with_N_thre1}.

The $\MSEp$ of LSDA increases with $\nusers$ in Fig.~\ref{fig:with_N_thre1} in a way that it is not predicted by our estimation~\eqref{eq:MSEthre}. This difference is due to the approximation taken in \eqref{eq:LLN_Rx} when deriving the MSE formula and is reduced as the number of rounds observed increases, as shown in Fig.~\ref{fig:with_N_thre2}.

Both C-LSDA and PMDA improve their result as the number of users increases. The initial estimation of PMDA improves with $\nusers$, which allows the attack to achieve better results also when $\rho$ grows (Fig.~\ref{fig:with_N_thre2}). 
However, we note that C-LSDA eventually outperforms PMDA when increasing only $\nusers$ or $\rho$. On the one hand, the advantage PMDA gains by increasing $\nusers$ does not improve constantly if $\rho$ is kept fixed, while the MSE of C-LSDA keeps decreasing with $\nusers$. Also, C-LSDA is asymptotically efficient while PMDA is not, which means that C-LSDA will eventually outperform PMDA when increasing $\rho$ regardless of the number of users in the system.

\subsubsection{Performance with respect to $\meanuniformi$ and $\uniformi{i}$}

\begin{figure*}[!t]
\centering
\subfloat[]{\includegraphics[width=2.8in]{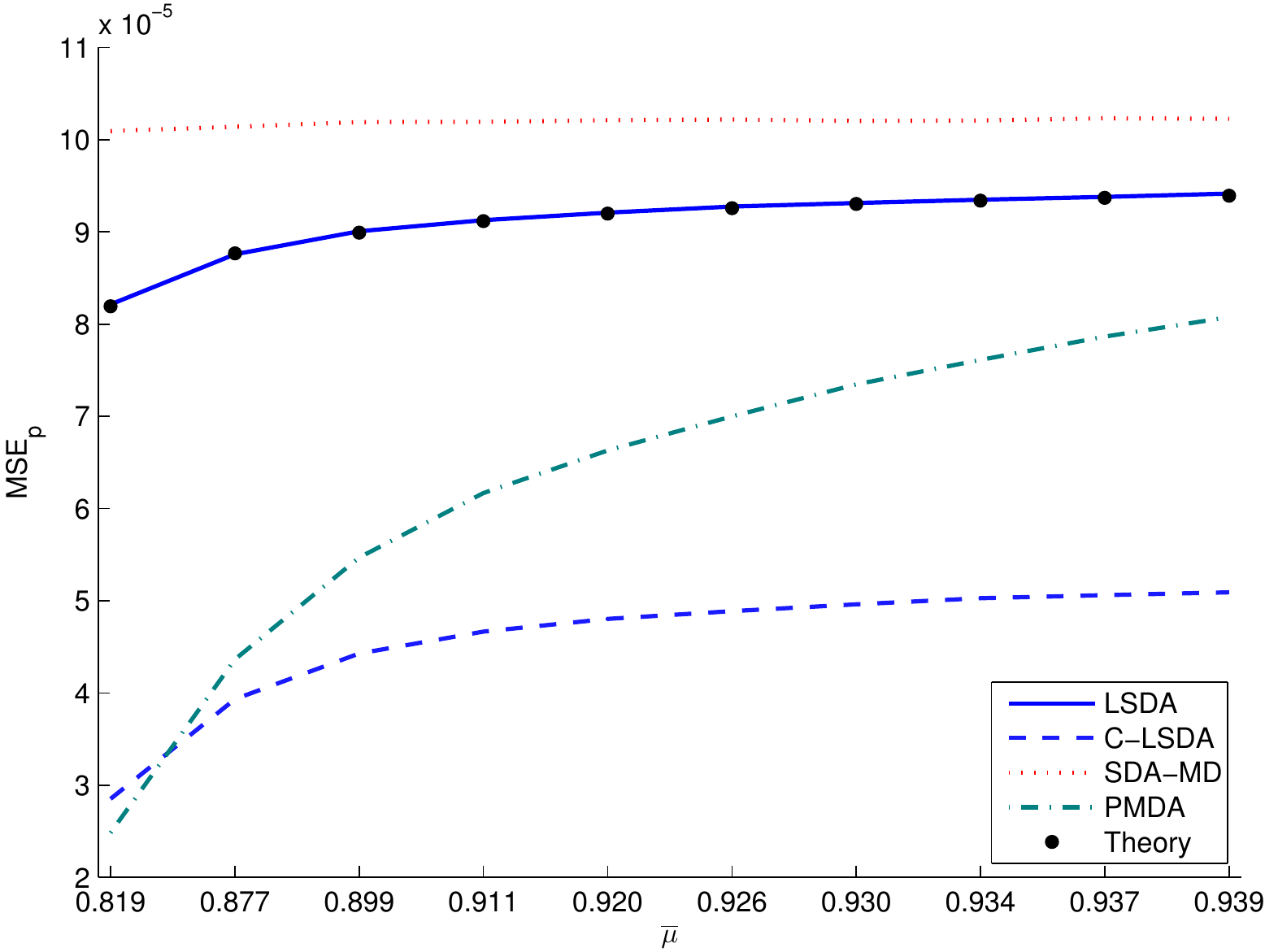} \label{fig:with_mu_thre1}}
\hfil
\subfloat[]{\includegraphics[width=2.8in]{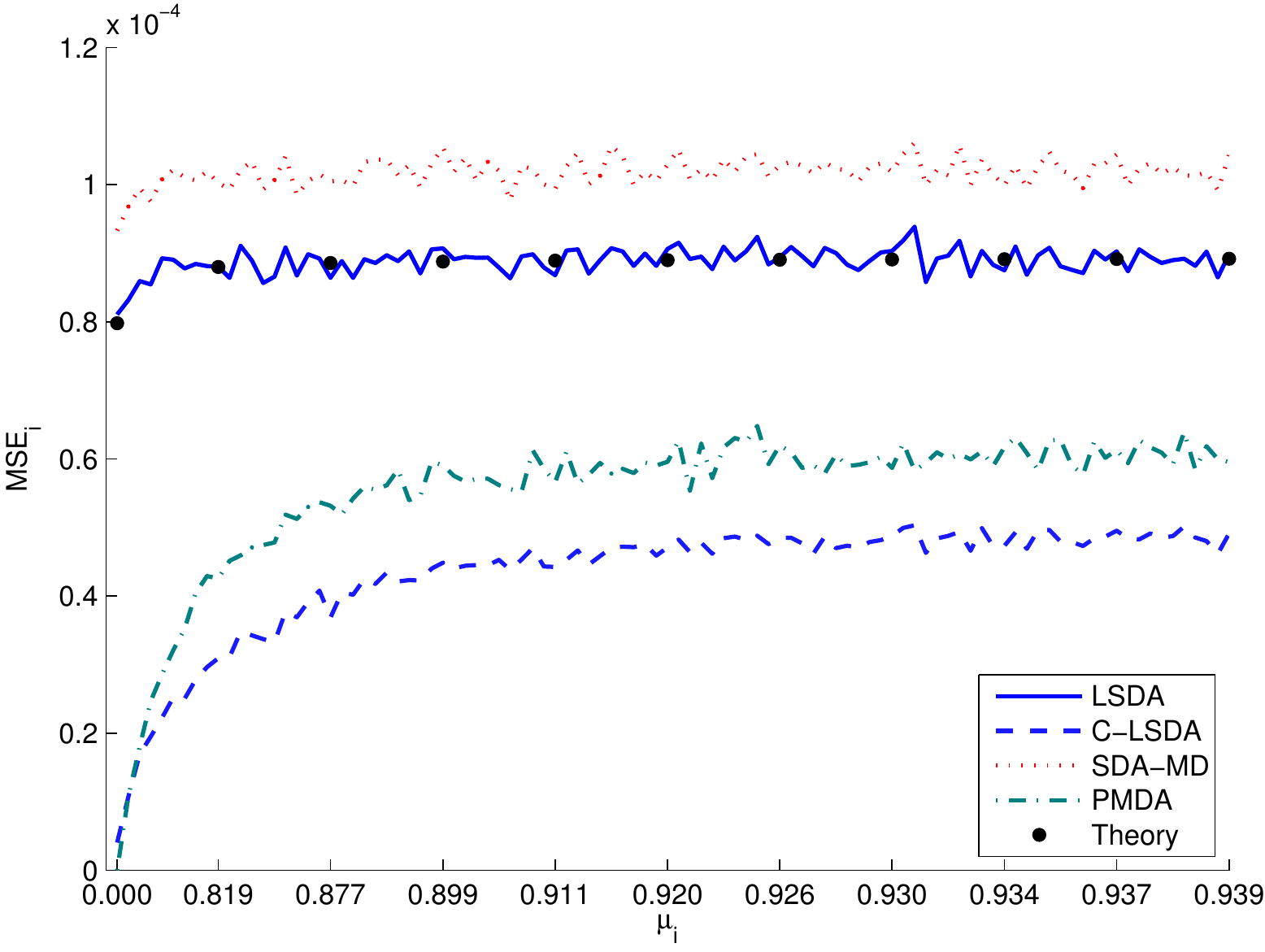} \label{fig:with_mu_thre2}}
\caption{$\MSEp$ evolution with $\meanuniformi$ in a scenario where $\uniformi{i}$ is constant (a) and $\MSEi$'s in an experiment where each user has a different value of $\uniformi{i}$ (b) ($\rho=10\,000$, $\nusers=100$, $\sendfreq{i}=1/\nusers$, $\thre=10$).}
\label{fig:with_mu_thre}
\end{figure*}

In this section, we first analyze the performance of the attacks with $\meanuniformi$ by increasing the number of friends $\nfriends$ assigned to all users in steps of $10$, as shown in Table~\ref{tab:parameters}. In this experiment, $\uniformi{i}=\meanuniformi$ for all $i$. The results are displayed in Fig.~\ref{fig:with_mu_thre1}. As we already hinted in Sect.~\ref{sec:performance}, user profiles are harder to estimate in LSDA (C-LSDA) when $\meanuniformi$ is closer to $1$, due to the sending behavior of the users becoming more random. PMDA is able to guess the correspondence between the input and output messages more often when users focus their traffic on few others, which happens at lower values of $\meanuniformi$. The results of SDA-MD, however, are almost independent of $\meanuniformi$ and $\uniformi{i}$\toremove{. This is because}{ since} \toremove{this attack}{it} groups the sending behavior of the users and thus is not able to exploit the hubness of the traffic, as explained in \cite{GlobalSIP}.

Figure~\ref{fig:with_mu_thre2} shows the MSE per sender profile, $\MSEi$, in a scenario where the parameters $\uniformi{i}$ are different for every user. The same reasons outlined above explain the performance of the attacks in this case.

\subsubsection{Performance with respect to the sending frequencies $\sendfreq{i}$}

\begin{figure}[!t]
\centering
  \includegraphics[width=2.8in]{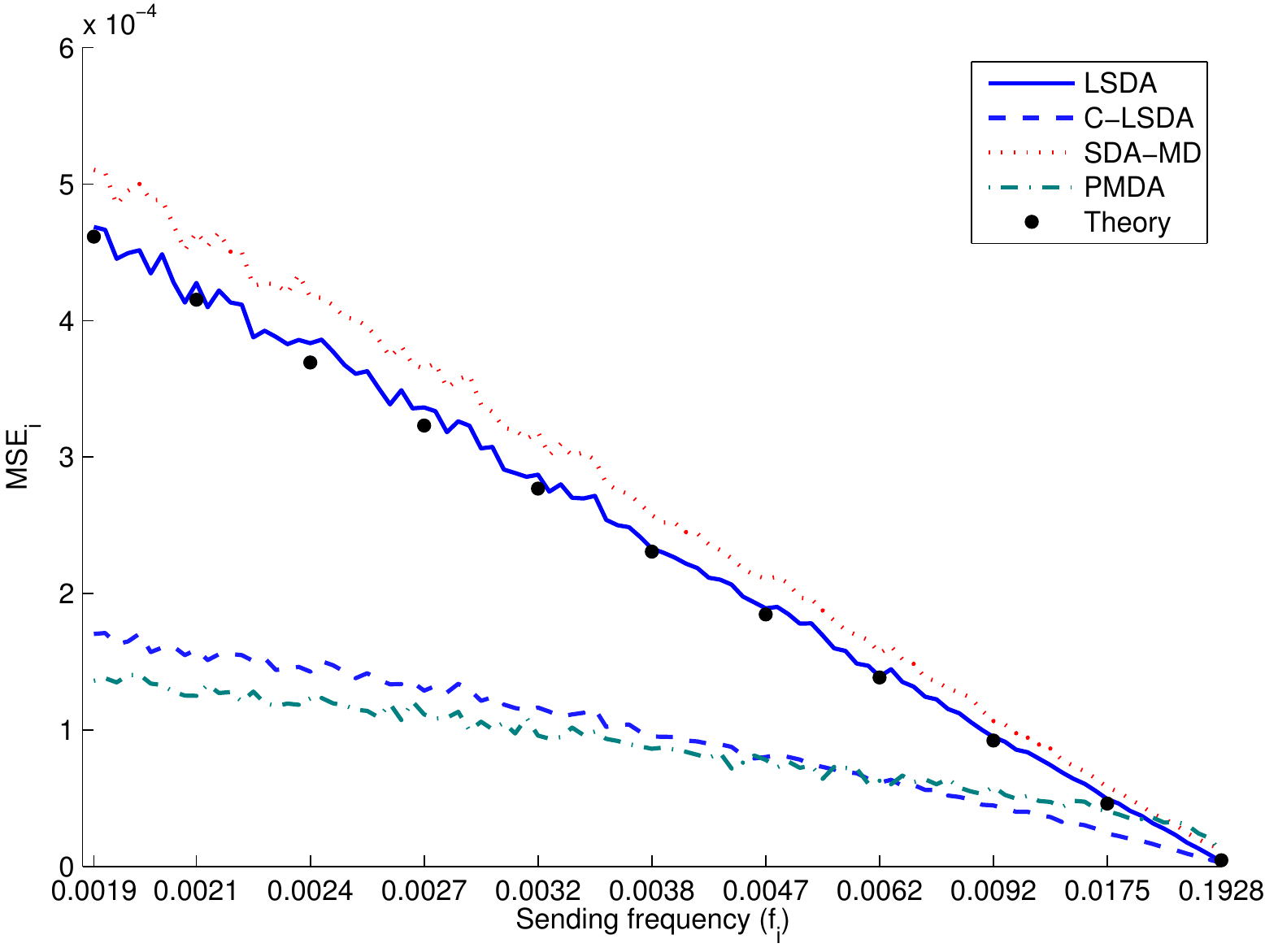}
\caption{Evolution of the MSE per sender profile with the sending frequency of the users $\sendfreq{i}$ ($\rho=10\,000$, $\nusers=100$, $\nfriends=25$, $\thre=10$).}
\label{fig:with_fi_thre}
\end{figure}

So far, we have assumed that every user participates in the system equally often. We now study what happens when this is not the case. Figure~\ref{fig:with_fi_thre} shows the $\MSEi$ in a scenario where $\sendfreq{i}$ varies among users. As we have outlined in Sect.~\ref{sec:performance}, the error in the estimation of a user's profile is tied to the participation of that user in the system. It is important to note that, for users who rarely send messages, PMDA seems to achieve better results than C-LSDA. This is because, for those users, the one-to-one correspondence of the messages sent and received in each round is very valuable information and LSDA does not exploit it. Nevertheless, if more round observations are available to the attacker, LSDA (C-LSDA) will eventually outperform PMDA.

\subsubsection{Performance with respect to the mix threshold $\thre$}

\begin{figure}[!t]
\centering
  \includegraphics[width=2.8in]{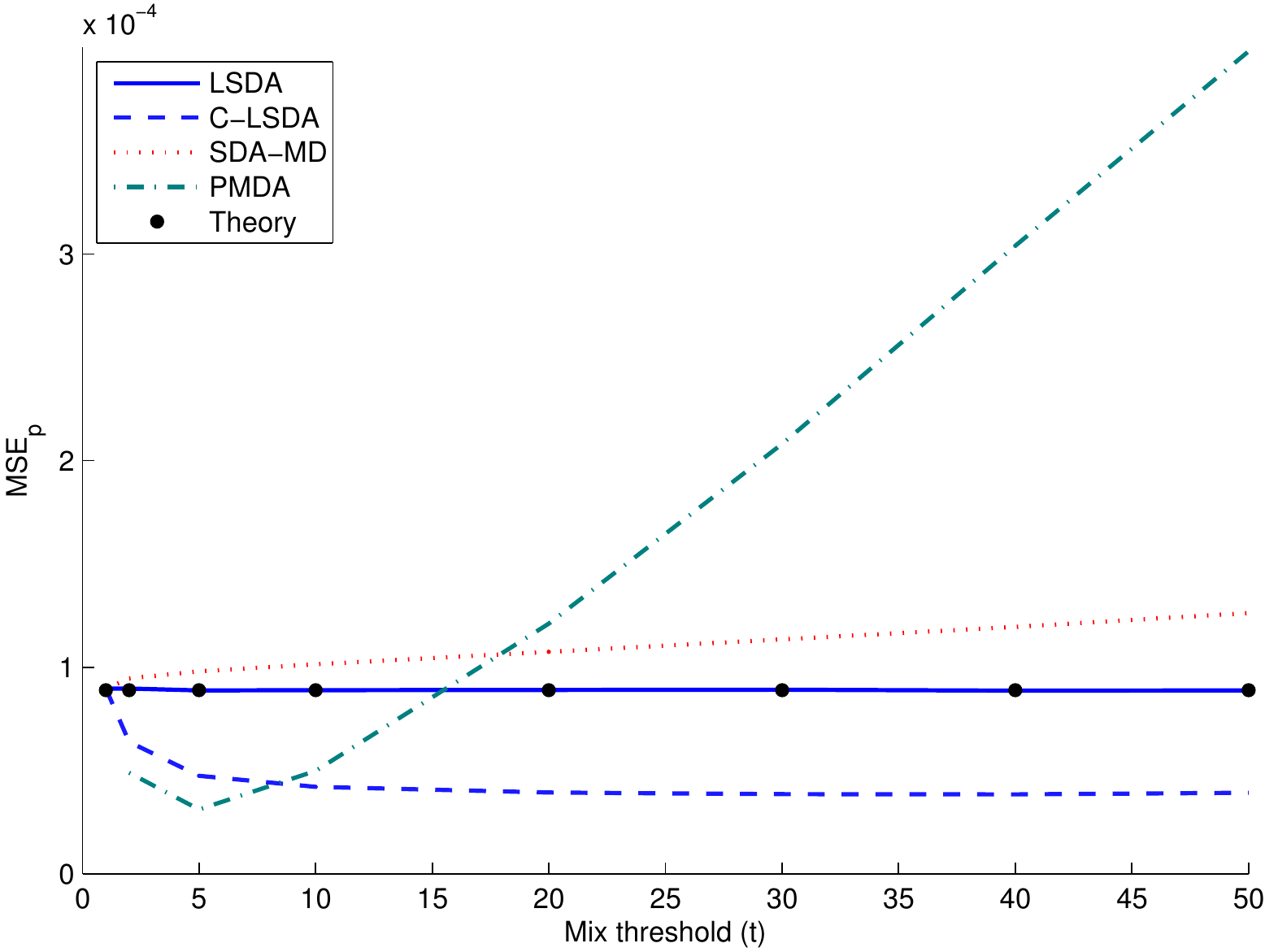}
\caption{$\MSEp$ evolution with the threshold $\thre$ ($\rho=10\,000$, $\nusers=100$, $\nfriends=25$, $\sendfreq{i}=1/\nusers$).}
\label{fig:with_t_thre}
\end{figure}

By observing \eqref{eq:MSEthre} one can see that the threshold $\thre$ of the mix has little influence on the $\MSEp$ of LSDA, becoming negligible as $\thre \gg 1$. This is reflected by our experiments, shown in Fig.~\ref{fig:with_t_thre}, where the error of LSDA soon becomes stable as the threshold of the mix grows. We must note that the time necessary to observe $\rho$ mixing rounds grows with the size of the threshold. Hence, although the error is constant with $\thre$, increasing the threshold delays the obtaining of accurate user profiles. Note that this protection the mix offers when increasing $\thre$ comes at the cost of an increased delay in the communications.

As expected, increasing the threshold has a negative effect on the other two attacks. The error in SDA-MD grows slightly with $\thre$, since increasing this parameter decreases the number of rounds where user $i$ does not participate and this makes harder to estimate the behavior of the messages not sent by $i$. On the other hand, as the threshold grows the number of plausible matchings increases and thus the likelihood that the most probable matching is the real one decreases. This in turn worsens PMDA's performance significantly.

\subsubsection{Comparison between attack principles}
Throughout the evaluation section we have studied three disclosure attacks that estimate users' profiles using statistics and optimization techniques. We now compare these attacks to Vida, the Bayesian inference-based machine learning algorithm proposed by Danezis and Troncoso in~\cite{DT09}. For the sake of comparison, we also test the efficacy of simply setting the negative probabilities output by the unconstrained LSDA and SDA-MD to zero (denoted as Z-LSDA and Z-SDA-MD, respectively). In order to illustrate the differences in computational load of the attacks, we have measured their running-time, carrying the experiment in a server with a Core2 Quad Q8300 $2.5$GHz processor, 8GB of RAM and Matlab version 7.13.0.564 (64-bit) on Ubuntu 12.04.3 LTS.

\begin{figure}[!t]
\centering
  \includegraphics[width=2.8in]{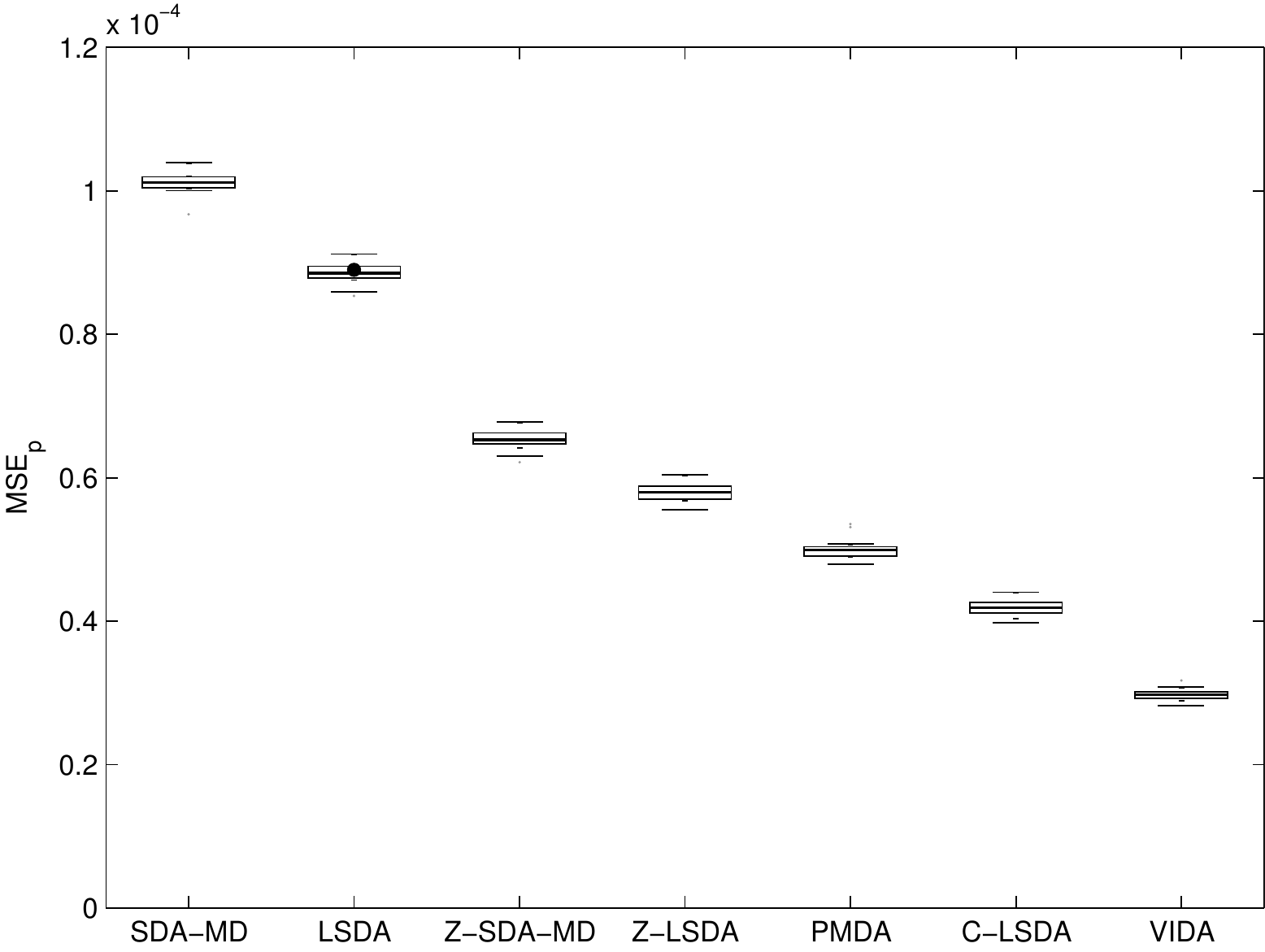}
\caption{Comparison between attack principles in a threshold mix ($\rho=10\,000$, $\nusers=100$, $\nfriends=25$, $\sendfreq{i}=1/\nusers$, $\thre=10$).}
\label{fig:boxplot_thre}
\end{figure}

Figure~\ref{fig:boxplot_thre} shows box plots representing the distribution of the $\MSEp$ obtained after 20 repetitions of our baseline experiment. We have already discussed that LSDA obtains an advantage over SDA-MD by solving the problem for all users simultaneously, but does not account for the one-to-one relationship between sent and received messages in the individual rounds of mixing as PMDA does. Recall, however, that both LSDA and SDA-MD outperform PMDA when the number of rounds observed is sufficiently large. When not considering the negative probabilities, both LSDA and SDA-MD improve their performance. However, only LSDA can be formulated as a constrained optimization problem (C-LSDA) and, when doing so, it achieves results really close to Vida's performance. This improvement comes at an increase in the computational cost: in this particular experiment, implementing LSDA as in \eqref{eq:compact_problem} and C-LSDA following \eqref{eq:landweber}, the latter was on average about $25$ times slower than LSDA (in each realization, C-LSDA took always less than $2$ seconds to finish).
The approach followed in Vida improves the profile estimations considerably with respect to the other attacks. While the effectiveness of Vida is desirable, it comes at a huge computational cost because each iteration of the algorithm requires finding a perfect matching in all the $\rho$ rounds observed. In this case, Vida was on average $280\,000$ times slower than LSDA (each realization took always more than $4$ hours to finish).

\subsection{Results: Pool mix}

We now proceed to evaluate LSDA's profiling performance when messages are anonymized using a threshold binomial pool mix. We recall that, in a threshold binomial pool mix, arriving messages are stored in a pool and leave the mix each round (i.e., when $\thre$ messages are received) with probability $\alpha$. Otherwise, messages stay in the pool until the next round, when they are mixed with the arriving fresh messages and again probabilistically selected to be fired or not. Additionally, we compare LSDA with SDA-MD, the most effective attack in the literature that has been applied to pool mixes, which we have implemented as explained in Sect.~\ref{sec:exp_setup}.

In this case, we only analyze the performance of the attacks with the firing probability $\alpha$ since, as we show in Appendix~\ref{sec:appendix_pool} with \eqref{eq:MSEpool}, the behavior of LSDA in the pool mix with all the other system parameters (i.e., $\rho$, $\nusers$, $\uniformi{i}$, $\sendfreq{i}$, $\thre$) does not change with respect to the threshold mix case.

\subsubsection{Performance with respect to the firing probability $\alpha$}

\begin{figure}[!t]
\centering
  \includegraphics[width=2.8in]{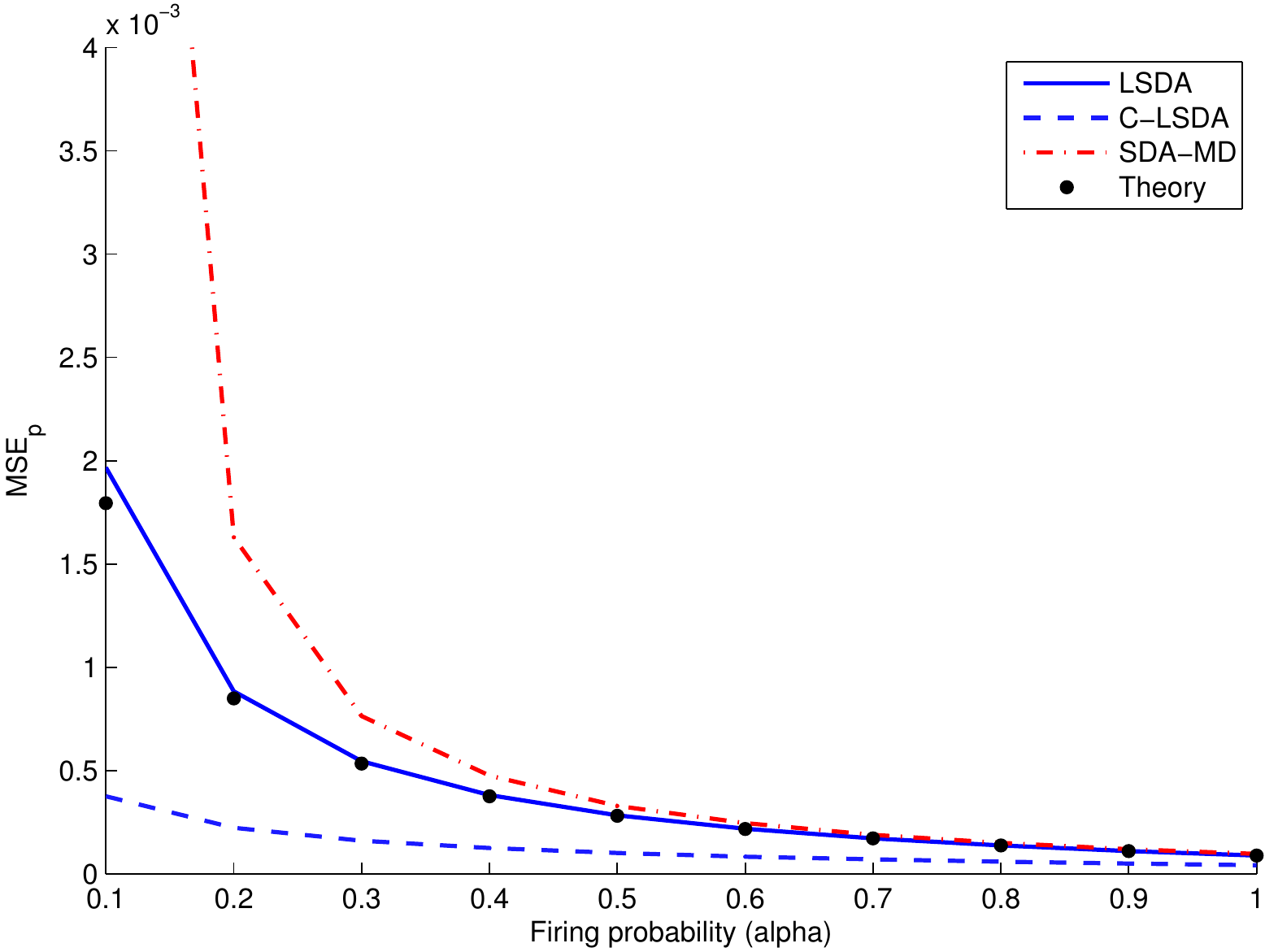}
\caption{$\MSEp$ evolution with the firing probability $\alpha$ ($\rho=10\,000$, $\nusers=100$, $\nfriends=25$, $\sendfreq{i}=1/\nusers$, $\thre=10$).}
\label{fig:MSE_pool_alpha}
\end{figure}

Given the operation of the mix, the delay (in rounds) suffered by messages traversing the mix follows a geometric distribution with parameter $\alpha$ and hence its mean is $(1-\alpha)/\alpha$. In Fig.~\ref{fig:MSE_pool_alpha}, we illustrate the tradeoff between the profiling accuracy of LSDA and $\alpha$. As expected, small values of $\alpha$ (i.e., large delays) result in larger error than when messages abandon the mix very fast. The longer the delay, the more messages participate in the mixing (\toremove{recall that }{}the mean size of the pool is $\frac{t-\alpha t}{\alpha}$), and the more difficult it is to estimate relations between senders and receivers. One can also see that the empirical error closely follows the prediction given by \eqref{eq:MSEpool}.

The figure also shows the evolution of the MSE per transition probability $\MSEp$ of Mathewson and Dingledine's SDA-MD. When the firing probability is low, messages from users are carried further to the next rounds. This, in time, reduces the number of rounds that can be used to estimate the behavior of all users but Alice (similarly to what happens when increasing $\thre$), making the estimation of the profiles increasingly difficult.

\subsubsection{Comparison between attack principles}
\begin{figure}[!t]
\centering
  \includegraphics[width=2.8in]{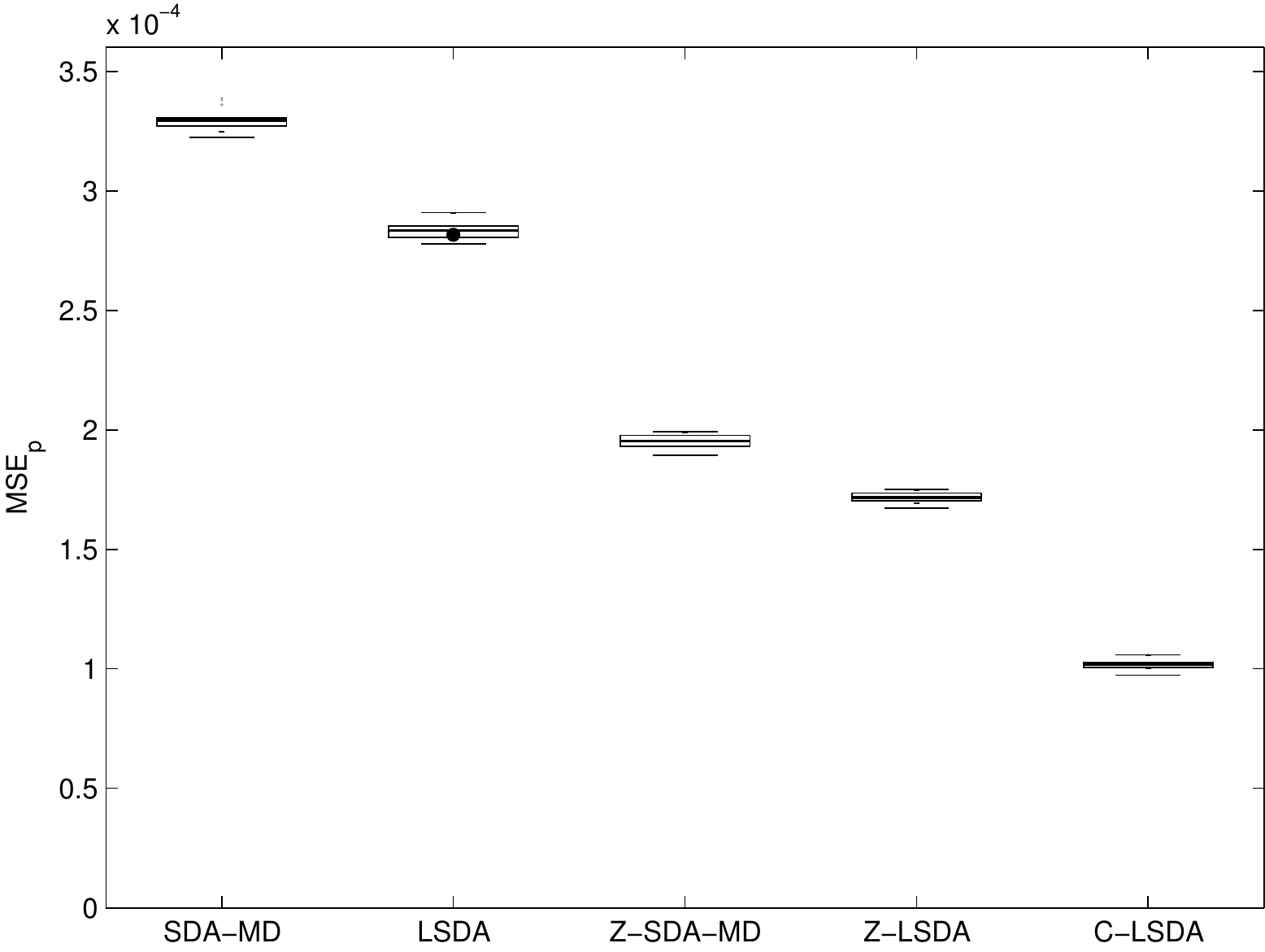}
\caption{Comparison between attack principles in a pool mix ($\rho=10\,000$, $\nusers=100$, $\nfriends=25$, $\sendfreq{i}=1/\nusers$, $\thre=10$, $\alpha=0.5$).}
\label{fig:boxplot_pool}
\end{figure}

We show, in Fig.~\ref{fig:boxplot_pool}, box plots representing the distribution of the $\MSEp$ for LSDA, C-LSDA, SDA-MD, Z-LSDA and Z-SDA-MD in a pool mix with $\alpha=0.5$. These results emphasize that LSDA (C-LSDA) is the attack which performs better in the pool mix.

\section{Extensions and Limitations of the Least Squares Disclosure Attack}
\label{sec:discussion}


\subsection{Real-world behavior}
In a realistic scenario, the frequency with which users send messages to the anonymous communication system, as well as their choices of recipients (i.e., their sending profiles), are likely to evolve over time. However, the analysis developed in this paper assumes that profiles and sending frequencies are static and hence it is only strictly applicable to observed time windows in which users' behavior does not change. As the length of the window increases, users' behavior evolution causes a deviation of the adversary's observation from these assumptions, and the accuracy of the error prediction diminishes accordingly. We now discuss how our analysis can be applied and extended to take into account behavior evolution, and empirically evaluate the performance of the static and non-static error estimation against real data.

Evolving behavior can be modeled by a stochastic process describing the collection of random variables that represent potential users' profiles. Hence, the behavior leading to a particular observation can be modeled by the random processes $\{F_i^r\}$ and $\{\Prob{j}{i}^r\}$ whose respective realizations $\{\sendfreq{i}^r\}$ and $\{\prob{j}{i}^r\}$ represent the actual values of the sending frequencies and transition probabilities in each round. We note that such a model can accommodate temporal changes of the profiles due to the evolution of users' individual preferences, as well as changes due to interactions between users (e.g., replies to messages or groups of receivers often contacted simultaneously).

In this scenario, LSDA can be used to estimate the average value of the random process $\{\Prob{j}{i}^r\}$ representing the users' \emph{average sending profiles} in the observed period, i.e., the average proportion of messages sent in this period to each of the possible recipients of the users. Notice that this corresponds to a frequentist interpretation of the user profiles. In fact, it can be shown that LSDA is an \emph{unbiased and efficient estimator} of the users' average sending profile as long as the sending profiles in each round $\prob{j}{i}^r$ can be modeled as a realization of a wide-sense stationary process $\{\Prob{j}{i}^r\}$ \cite{techrep}.

Following the approach carried out in the Appendices for the static case, it is possible to characterize the estimation error for the case of time-varying profiles \cite{techrep}. This analysis reveals that time-varying profiles increase the variance of the outputs given the input observations, which in turn slightly increases the adversary's estimation error. On the other hand, time-varying sending frequencies increase the variance of the input process, providing the attacker with a wider variety of input observations than in the static case. This in turn reduces LSDA's MSE and allows the adversary to obtain better estimates of the user profiles \cite{techrep}.



In order to show the usefulness of our methodology in presence of real traffic, we evaluate the performance of our error estimation analysis using the public Enron dataset.\footnote{\url{http://www.cs.cmu.edu/~./enron/}} This dataset consists of $N=294$ senders from this database (we have removed 11 users that send less than 20 messages during the collection period), which send a total of $220\,032$ messages to $17\,009$ receivers. In this experiment, messages are used as input to an anonymous channel implemented as a threshold mix with threshold $t=10$, obtaining $22\,000$ mixing rounds. Note that, although the model in Sect.~\ref{sec:sysmodel} assumes that the number of senders and receivers is the same, it is straightforward to extend the attack to the case where these numbers do not coincide by redefining the sizes of the involved vectors and matrices. Since we do not have access to the real profiles and sending frequencies of the users that generated the trace, for our experiments we compute the real probabilities $\prob{j}{i}$ as the proportion of messages from user $i$ sent to $j$ in the observed period, and the parameters $\sendfreq{i}$ as the fraction of all incoming messages to the mix that were sent by user $i$.

\begin{figure}[!t]
\centering
  \includegraphics[width=2.8in]{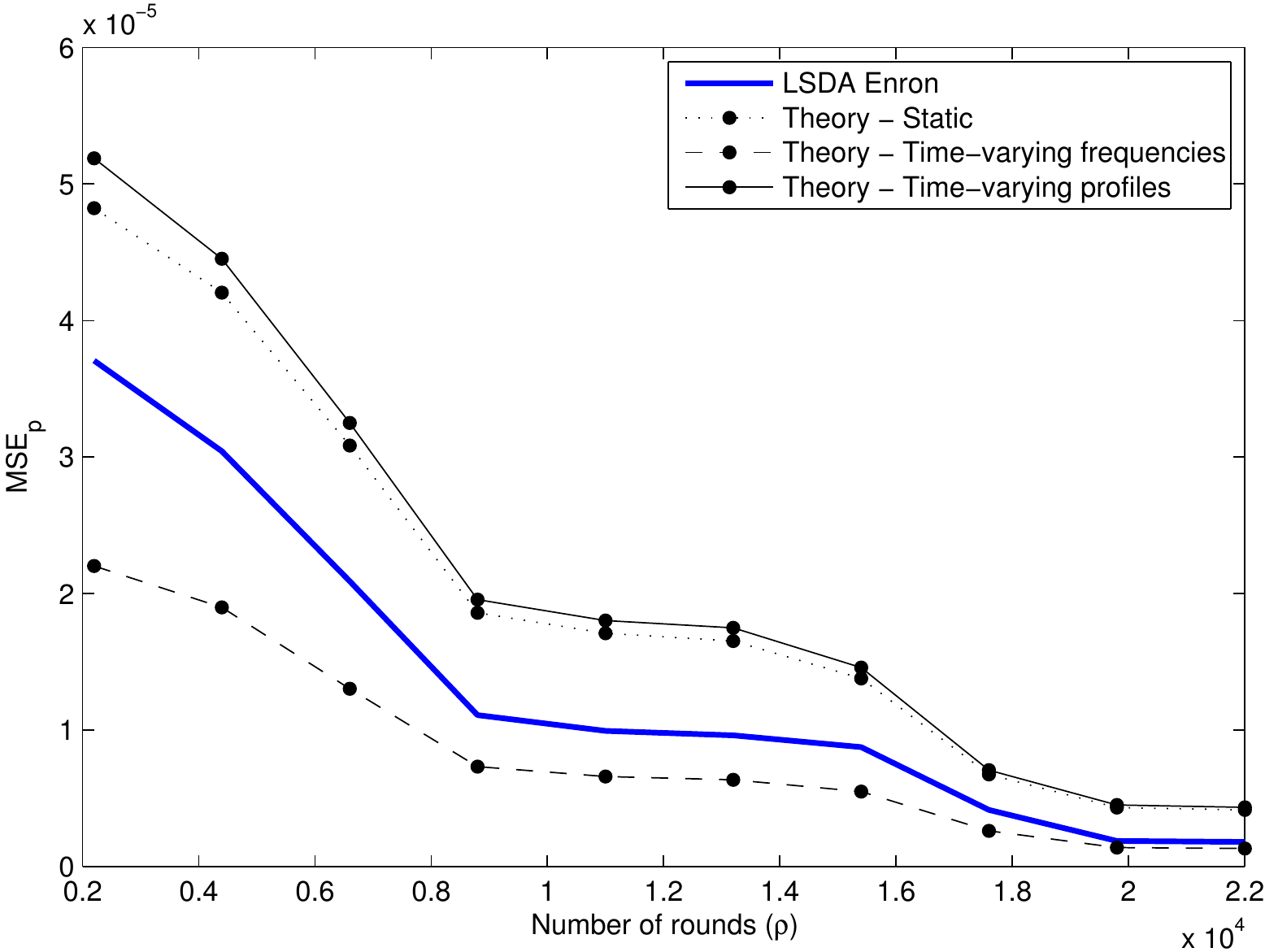}
\caption{Empirical and theoretical $\MSEp$ of LSDA against the Enron dataset.}
\label{fig:MSE_ENRON}
\end{figure}

We estimate the sender profiles of the users that participate in the system using LSDA for $\rho=\{2\,200,4\,400,\ldots,22\,000\}$ rounds, and compute the MSE for each estimated transition probability $\pest{j}{i}$. We plot in Fig.~\ref{fig:MSE_ENRON} the average MSE per transition probability ($\MSEp$) of $100$ users (thick straight line) that send messages throughout the full observation period. The figure also shows the theoretical estimation of the error using \eqref{eq:MSEthre} (dotted line), and using the error prediction for non-static cases provided in \cite{techrep} for the cases when sending frequencies vary over time while profiles stay invariant (dashed lines), and when profiles vary over time while sending frequencies remain fixed (thin straight line).

As expected, real traffic does not behave according to the assumptions made in the error estimation performance analysis, and hence the results are less accurate than in the simulations in Sect. \ref{sec:eval}. Yet, the predictor based on static parameters is not far from the empirical result (the predicted transition probabilities are off by at most $1.5 \times 10^{-5}$), and correctly follows the trend of the error as more information is made available to the adversary. In fact, in \cite{techrep} we show that our formula in \eqref{eq:MSEthre} practically gives an upper bound to the empirical MSE when the number of users in the system is relatively large. Therefore, the results provided in this paper conversely serve as a lower bound to the {\em privacy loss} of mixes as rounds of observations become available to the attacker, even when users' profiles and sending frequencies evolve with time. This highlights the usefulness of the theoretical results provided in this paper, pointing at a fundamental weakness of existing mix-based anonymous communication systems. Such results generalize the findings of Kesdogan et al~\cite{KAP02}, proving that anonymity protection limits are caused by diversification in user behavior rather than by the observation of changing anonymity sets.

As a final remark, note that LSDA can be used to infer how fast the sending behavior of the users evolves with time. To this end, the adversary can split the observation in shorter time windows (which may overlap) and use LSDA to obtain the users' average sending profile within each subset of consecutive rounds of mixing. By varying the size of the window the adversary can detect when the profiles change.

\subsection{Adversarial prior knowledge}
In some cases it might be possible that the adversary has some prior information on the transition probabilities. While it is possible to modify the machine learning approach~\cite{DT09} to account for this extra knowledge, this is non-trivial for SDA or PMDA. In contrast, the least squares formulation can be easily adapted to consider this additional information: if the values of some probabilities are known, the attack can be extended in a similar way as we did to show that the original SDA is a particular, but largely suboptimal, instance of LSDA. On the other hand, the prior knowledge may be given as a set of constraints on the transition probabilities (e.g., the adversary knows that user $i$ contacts user $j$ at least half of the times). In that case, our iterative implementation of C-LSDA in Sect.~\ref{sec:clsda} could be adapted to such scenario by just projecting the solution onto the new set of constraints.

\subsection{User and background cover traffic}
Cover traffic, i.e., fake messages generated by the users and/or the mix, makes harder for the attacker to infer the users' sending profiles~\cite{MW07}. In this scenario, dummy messages can be regarded as noise in the input and output observations. In order to derive the LSDA estimator in this case, we would look for the profiles which minimize the error between the observed and the expected output, now considering that this output consists of real and dummy messages and that the input observations are noisy.

\section{Conclusions}
\label{sec:conclusion}

We have introduced the Least Squares Disclosure Attack, which estimates user profiles by minimizing the prediction error of the output given the input. By modeling the estimation of profiles as a least squares problem, we are able to obtain analytic results that predict the profiling error for a given set of system parameters. This prediction is very accurate when users' behavior is static, and when this assumption does not hold it gives a good approximation that correctly follows the trend of the error as more information is made available. This feature permits the designer of a high-latency anonymous communication system to choose parameters that provide a desired level of protection depending on the population characteristics without the need to perform expensive simulations~\cite{DT09,TGPV08}.

Under the hypotheses of static users' behavior, we have proven that, contrary to other approaches~\cite{TGPV08}, this least squares estimator is asymptotically efficient, meaning that if the attacker is given enough observations of the system, the error on the estimation of the user profiles will approach zero. This finding generalizes previous results on the limits of the protection of mix-based anonymous communication systems, proving that such limits are imposed by varying user behavior rather than changing anonymity sets.

Our experiments confirm that, in simple threshold mixes, LSDA outperforms the state-of-the-art version of the Statistical Disclosure Attack~\cite{MD04} and that a more sophisticated implementation of LSDA gives results close to those of the computationally expensive Bayesian inference approach~\cite{DT09} at an affordable cost. Our attack is not limited to the analysis of threshold mixes but can be easily extended to more complex mixing strategies such as pool mixes~\cite{WIFS12}. In these complex scenarios, the Bayesian approach is untractable and LSDA yields the best results.


%

\appendices

\section{Derivation of $\MSEi$ for the threshold mix} 
\label{sec:appendix_thre}

Our goal is to derive an expression for the Mean Squared Error per user, $\MSEi\doteq\sum_{j=1}^{\nusers} \text{E}\left\{ \left(\prob{j}{i}-\probest{j}{i}\right)^2 \right\}$, when using the LSDA estimator in \eqref{eq:LSDAthre} in a threshold mix scenario. To this end, we first remark that the unconstrained least squares estimate \eqref{eq:LSDAthre} is unbiased: it is straightforward to show that $\text{E}\{\recvprofest{j}\}=\recvprof{j}$ by using the law of total expectation and $\text{E}\left\{ \recvvr{j}{} | \sendm \right\}=\sendm \cdot \recvprof{j}$:
\begin{equation} \label{eq:unbiased}
\begin{array}{lcl}
 \text{E}\{ \recvprofest{j} \}&=&\text{E}\left\{ \text{E}\left\{ \recvprofest{j} | \sendm{} \right\} \right\}
 = \text{E}\left\{ \left( \sendm^T \sendm \right)^{-1} \sendm^T \text{E}\left\{ \recvvr{j}{} | \sendm\right\} \right\}\\
 &=& \text{E}\left\{ \left( \sendm^T \sendm \right)^{-1} \sendm^T \sendm \cdot \recvprof{j} \right\}
 = \recvprof{j}\,.
\end{array}
\end{equation}

On the other hand, we can use the law of total variance together with the fact that, from \eqref{eq:unbiased}, $\text{Var}\{\mbox{E}\{\probest{j}{i}|\sendm\}\}=0$ and $\text{Cov}\{\text{E}\{\probest{j}{i}|\sendm\},\text{E}\{\probest{j}{k}|\sendm\}\}=0$ for $k\neq i$, to express the covariance matrix of $\recvprof{j}$ as
\begin{equation} \label{eq:covpj}
 \bt \Sigma_{\recvprof{j}} = \text{E}\{\bt \Sigma_{\recvprof{j}|\sendm}\}
                           = \text{E}\{(\sendm^T\sendm)^{-1}\sendm^T \bt \Sigma_{\recvvr{j}{}|\sendm} \sendm(\sendm^T\sendm)^{-1}\}
\end{equation}
where $\bt \Sigma_{\recvvr{j}{}|\sendm}=\text{E}\{(\recvvr{j}{}-\text{E}\left\{\recvvr{j}{}|\sendm\})(\recvvr{j}{}-\text{E}\{\recvvr{j}{}|\sendm\})^T|\sendm\right\}$.

We model $\{\sendr{1}{r}, \cdots, \sendr{\nusers}{r}\}$ jointly as a multinomial distribution with $t$ trials and probabilities $\{\sendfreq{1}, \cdots, \sendfreq{\nusers}\}$. Also, we note that $\sendr{i}{r}$ and $\sendr{k}{s}$ are independent when $r\neq s$. In order to compute \eqref{eq:covpj}, we first point out that, since the input process is stationary and memoryless, then using the Law of Large Numbers we can write
\begin{equation} \label{eq:LLN_Rx}
 \lim_{\rho \rightarrow \infty} \sendm^T \sendm / \rho \rightarrow \autocorr
\end{equation}
where $\autocorr$ is the autocorrelation matrix of the input process. We can write this autocorrelation matrix, using
\begin{eqnarray}
 \text{E}\{ \sendr{i}{2} \}&=&\left(\thre^2-\thre\right)\sendfreq{i}^2 + \thre \sendfreq{i}\label{eq:X2} \\
 \text{E}\{ \sendr{i}{}\sendr{k}{} \}&=&\left(\thre^2-\thre\right) \sendfreq{i} \sendfreq{k}\,\qquad\text{ for } i\neq k \label{eq:XX}
\end{eqnarray}
as
\begin{equation} \label{eq:Rx}
 \autocorr=\thre\left[\sendfreqm + \left(\thre-1\right)\f \cdot \f^T\right]
\end{equation}
where $\f\doteq[f_1,\cdots,f_N]^T$ and $\sendfreqm\doteq\text{diag}\{\f\}$. Applying the Sherman-Morrison formula \cite{shermanmorrison}, the inverse of this autocorrelation matrix can be written as
\begin{equation} \label{eq:invrx}
 \autocorr^{-1}=\frac{1}{t}\left[\sendfreqm^{-1}-\left(1-\frac{1}{\thre}\right)\mathbf{1}_{\nusers\times \nusers} \right]\,.
\end{equation}

Assuming that the number of observed rounds $\rho$ is large, we can approximate \eqref{eq:covpj} as
\begin{equation} \label{eq:covpj2}
 \bt \Sigma_{\recvprof{j}} \approx \frac{1}{\rho^2} \autocorr^{-1} \text{E}\{ \sendm^T \bt \Sigma_{\recvvr{j}{}|\sendm} \sendm \} \autocorr^{-1}\,.
\end{equation}

We now focus on the term $\text{E}\{ \sendm^T \bt \Sigma_{\recvvr{j}{}|\sendm} \sendm\}$. We model $\recvr{j}{r}|\sendm{}$ as the sum of $\nusers$ binomial processes with $\send{i}{r}$ trials and probabilities $\prob{j}{i}$, for $i=1, 2, \cdots, \nusers$. Note that $\recvr{j}{r}$ and $\recvr{j}{s}$ are independent for $r\neq s$. Let $\binvar{i}\doteq\prob{j}{i}\cdot(1-\prob{j}{i})$ and $\binvarm\doteq\text{diag}\{\binvar{1}, \cdots, \binvar{\nusers}\}$. Then, $\mathbf{\Sigma}_{\recvvr{j}{}|\sendm{}}$ is a diagonal matrix whose $(r,r)$-th element is
\begin{equation}
 \left(\mathbf{\Sigma}_{\recvvr{j}{}|\sendm{}}\right)_{r,r}=\sum_{i=1}^{\nusers} \send{i}{r} \binvar{i}\,.
\end{equation}

Operating, we get that the $(m,n)$-th element of $\text{E}\{ \sendm^T \bt \Sigma_{\recvvr{j}{}|\sendm} \sendm\}$ is
\begin{equation}
 \left(\text{E}\{ \sendm^T \bt \Sigma_{\recvvr{j}{}|\sendm} \sendm\} \right)_{m,n}=\rho \sum_{i=1}^{\nusers} \binvar{i} \cdot \text{E}\{ \sendr{i}{} \sendr{m}{} \sendr{n}{} \}
\end{equation}
and therefore we can write this term as
\begin{equation} \label{eq:midterm}
\begin{array}{l}
 \text{E}\{ \sendm^T \mathbf{\Sigma}_{\recvvr{j}{}|\sendm} \sendm\}=\vspace{0.1cm} \\ 
 \rho \left[ \sendfreqm \left( \eta_j \thre^{(3)} \mathbf{1}_{\nusers\times\nusers} + \binvarm \mathbf{1}_{\nusers\times\nusers} \thre^{(2)} 
 + \mathbf{1}_{\nusers\times\nusers} \binvarm \thre^{(2)} \right) \sendfreqm \right] \vspace{0.1cm} \\
 + \ \rho \left[\left(\eta_j \thre^{(2)} \mathbf{I}_{\nusers\times\nusers} + \thre \binvarm \right) \sendfreqm \right]
 \end{array}
\end{equation}
where $\eta_j\doteq\sum_{i=1}^{\nusers} \sendfreq{i} \binvar{i}$ and $\thre^{(n)}\doteq\thre\cdot(\thre-1) \cdots (\thre-n+1)$.

Plugging \eqref{eq:midterm} into \eqref{eq:covpj2} we get an approximation of $\mathbf{\Sigma}_{\recvprof{j}}$. Now, taking each of the diagonal elements of this matrix, which are $\text{Var}\{\probest{j}{i}\}$ for $i=1, \cdots, \nusers$ and adding them along $j$ to obtain $\MSEi\doteq\sum_{j=1}^{\nusers} \text{Var}\{ \probest{j}{i} \}$, we finally get
\begin{equation}
 \MSEi\approx \frac{1}{\rho} \left\{ \left(\sendfreq{i}^{-1}-1\right)\left(1-\frac{1}{\thre}\right)\meanuniformi + \frac{\sendfreq{i}^{-1}}{\thre} \cdot \uniformi{i}\right\}\,.
\end{equation}

\section{Derivation of $\MSEi$ for the pool mix}
\label{sec:appendix_pool}

We aim here at deriving an expression for the MSE in the estimation of the sender profile of user $i$, previously defined as $\MSEi\doteq\sum_{i=1}^{\nusers} \text{E}\{ \left( \prob{j}{i}-\probest{j}{i} \right)^2 \}$, for the LSDA estimator in the pool mix \eqref{eq:LSDApool}. To this end, we follow the same approach as for the threshold mix. We will assume that $\Nzero=\bt 0$, as the impact of the initial conditions can be neglected for large $\rho$.

We start by showing that this estimator is unbiased. \toremove{Recall first that, f}{F}ollowing \eqref{eq:expZpool}, we can write $\text{E}\{ \recvvr{j}{}|\sendm\}=\sendpoolmest \cdot \recvprof{j}$. Then,
\begin{eqnarray*}
  \text{E}\{ \recvprofest{j} \}&=&\text{E}\left\{ \text{E}\left\{ \recvprofest{j} | \sendm \right\} \right\}
  = \text{E}\left\{ \left( \sendpoolmest^T \sendpoolmest \right)^{-1} \sendpoolmest^T \text{E}\left\{ \recvvr{j}{} | \sendm\right\} \right\} \\
  &=& \text{E}\left\{ \left( \sendpoolmest^T \sendpoolmest \right)^{-1} \sendpoolmest^T \sendpoolmest \cdot \recvprof{j} \right\} 
  =\recvprof{j}\,.
\end{eqnarray*}

Using the law of total variance, we can now write
\begin{equation} \label{eq:covpj_pool}
  \bt \Sigma_{\recvprof{j}} = \text{E}\{(\sendpoolmest^T\sendpoolmest)^{-1}\sendpoolmest^T \bt \Sigma_{\recvvr{j}{}|\sendm} \sendpoolmest(\sendpoolmest^T\sendpoolmest)^{-1}\}
\end{equation}
where $\sendpoolmest = \B \sendm$. As in the threshold mix case, we approximate \eqref{eq:covpj_pool}, assuming that the number of observed rounds $\rho$ is large enough, as
\begin{equation} \label{eq:covpj2_pool}
 \bt \Sigma_{\recvprof{j}} \approx \frac{1}{\rho^2} \autocorrpool^{-1} \text{E}\{\sendpoolmest^T \bt \Sigma_{\recvvr{j}{}|\sendm} \sendpoolmest\} \autocorrpool^{-1}\,.
\end{equation}
The $(m,n)$-th element of the autocorrelation matrix in the pool mix case, $\autocorrpool$, is 
\begin{equation}
 \left( \autocorrpool \right)_{m,n}=\frac{1}{\rho} \sum_{k=1}^{\rho} \sum_{r=1}^k \sum_{s=1}^k \text{E}\{ \sendr{m}{r} \sendr{n}{s} \} \alpha^2 (1-\alpha)^{2k-r-s}\,.
\end{equation}
In order to get an expression for this matrix, we can use \eqref{eq:X2} and \eqref{eq:XX} since the distribution of the input process in the pool mix is the same as in the threshold mix. If we assume that $\rho \gg 1/\alpha$, and define $\alphaq=\alpha/(2-\alpha)$, then we can approximate this autocorrelation matrix by $\autocorrpool \approx \alphaq \thre  \sendfreqm  + \left(\thre^2 - \alphaq \thre\right) \f \cdot \f^T$, 
whose inverse is
\begin{equation} \label{eq:autocorrinv_pool}
 \autocorrpool^{-1} \approx \frac{1}{\rho \alphaq \thre}\left[ \sendfreqm^{-1} - \left( 1 - \frac{\alphaq}{\thre}\right) \bt 1_{\nusers \times \nusers}\right]\,.
\end{equation}

We now focus on $\text{E}\{ \sendpoolmest^T \bt \Sigma_{\recvvr{j}{}|\sendm} \sendpoolmest \}$ in \eqref{eq:covpj2_pool}. In this case, the random variables $\recvr{j}{r}$ in different rounds are not independent and therefore we cannot use the $\bt \Sigma_{\recvvr{j}{}|\sendm}$ that we derived for the threshold mix scenario. Using the law of total variance, it can be shown that
\begin{equation*}
\begin{array}{ll}
 \text{Var}\{\recvr{j}{r}|\sendm\}=\sum\limits_{k=1}^{r} \sum\limits_{i=1}^{\nusers} \send{i}{k}&\bigl(\, \prob{j}{i} \alpha(1-\alpha)^{r-k}\\
											      &-\prob{j}{i}^2\alpha^2(1-\alpha)^{2(r-k)} \,\,\bigr) \\
\end{array}
\end{equation*}
\begin{equation}
\begin{array}{ll}
 \text{Cov}\{\recvr{j}{r}, \recvr{j}{s}|\sendm\}=-\sum\limits_{k=1}^{\text{min}(r,s)} \sum\limits_{i=1}^{\nusers} \send{i}{k} \prob{j}{i}^2 \alpha^2 (1-\alpha)^{r+s-2k}
\end{array}
\end{equation}
which equals, in matricial form, to
\begin{equation} \label{eq:covyj_pool}
 \bt \Sigma_{\recvvr{j}{}|\sendm}=\text{diag}\{ \B \sendm \recvprofm{j} \bt 1_{\nusers}\}-\B \cdot \text{diag}\{ \sendm \recvprofm{j}^2 \bt 1_{\nusers} \} \cdot \B^T
\end{equation}
where $\recvprofm{j}=\text{diag}\{ \recvprof{j} \}$.

The following steps are similar to those performed in the threshold mix case and consist only of laborious matrix multiplications. We omit the full description of these steps for practicality issues and outline the remaining process: using \eqref{eq:covyj_pool}, we compute $\text{E}\{ \sendpoolmest^T \bt \Sigma_{\recvvr{j}{}|\sendm} \sendpoolmest \}$. We multiply the resulting matrix left and right by \eqref{eq:autocorrinv_pool}, and then take the diagonal elements of the resulting matrix, which are an approximation of $\text{Var}\{\probest{j}{i}\}$. Adding these elements along $j$ and defining $\alpha_r\doteq\alpha(2-\alpha)/(2-\alpha(2-\alpha))$, we finally obtain
\begin{equation} \Fitcol{
\MSEi \approx \displaystyle\frac{1}{\rho}\left\{ (\sendfreq{i}^{-1}-1)\left[ \meanuniformi \left(\frac{1}{\alpha_r}-\frac{1}{\thre}\right) + \left(\frac{1}{\alphaq} - \frac{1}{\alpha_r}\right)\right]+\frac{\sendfreq{i}^{-1}}{\thre}\cdot \uniformi{i} \right\}}
\end{equation}

\bibliographystyle{IEEEtran}
\bibliography{IEEEabrv,references}

\end{document}